%% file: Higgs2000.tex
\documentclass[12pt,a4paper]{article}
\usepackage{epsfig}
\usepackage{feynmf}
\input{definitions}

\begin{document}\setlength{\unitlength}{1mm}
%
%
%
\begin{titlepage}
\begin{center}
\begin{large}
EUROPEAN ORGANIZATION FOR NUCLEAR RESEARCH (CERN)
\end{large}
\end{center}
\begin{flushright}
CERN-EP/2000-138\\
November 13, 2000 \\
\end{flushright}
\setlength{\topmargin}{0.5cm}
\setlength{\oddsidemargin}{-0.2cm}
\vspace{1cm}
\begin{center}
\begin{LARGE}
{Observation of an Excess in the Search}
\end{LARGE}
\vskip 0.5cm
\begin{LARGE}
{for the Standard Model Higgs Boson at ALEPH}
\end{LARGE}
\end{center}
\vskip 1cm
\begin{center}
\begin{large}
{The {\sc ALEPH} Collaboration $^{*)}$}
\end{large}
\end{center}
\vspace{1cm}
\thispagestyle{empty}

\input{abstract}

\vspace{0.7cm}
\begin{center}
{\it (Submitted to Physics Letters B)}

\vspace{2.0cm}

$^*$) See next pages for the list of authors.
\end{center}
\end{titlepage}

\newpage
\pagestyle{plain}
\setcounter{page}{1}
%
%
\pagenumbering{arabic}
\normalsize
\setlength{\textheight}{23cm}
\setlength{\textwidth}{17cm}
\unitlength 1mm
\setlength{\topmargin}{-1cm}
\setlength{\oddsidemargin}{-0.5cm}
%
%
\input{authb}

\setcounter{page}{1}

\input{introduction}

\input{analyses}

\input{cl_results2}

\input{systematics}

\input{dissection}

\input{candidates}

\input{conclusions}

\section*{Acknowledgements}
We congratulate our colleagues from the accelerator divisions
for the very successful running of \LEP\ at high energies.
Without the extraordinary achievement of operating LEP at
energies much above the design value, these observations would
not have been possible.
We are
indebted to the engineers and technicians in all our institutions for
their contribution to the excellent performance of \ALEPH.  Those of
us from non-member countries thank \CERN\ for its hospitality.

\input{biblio.tex}
\end{document}

%% file: definitions.tex
\def\invpb{\ensuremath{{{\rm pb}^{-1}}}}

\def\Gc{\ensuremath{{\mathrm{GeV}/c}}}
\def\Gcs{\ensuremath{{\mathrm{GeV}/c^2}}}

\def\H{\ensuremath{{\mathrm H}}}
\def\h{\ensuremath{{\mathrm H}}}

\def\Z{\ensuremath{{\mathrm Z}}}
\def\W{\ensuremath{{\mathrm W}}}
\def\e{\ensuremath{{\mathrm e}}}
\def\l{\ensuremath{{\ell}}}

\def\WW{\ensuremath{{\mathrm W}^{+}{\mathrm W}^{-}}}

\def\ee{\ensuremath{{\mathrm e}^{+}{\mathrm e}^{-}}}
\def\epem{\ee}
\def\ffbar{\ensuremath{{\mathrm f}\bar{\mathrm f}}}
\def\bb{\ensuremath{{\mathrm b}\bar{\mathrm b}}}
\def\bbbar{\bb}
\def\qq{\ensuremath{{\mathrm q}\bar{\mathrm q}}}
\def\qqbar{\qq}
\def\qqg{\ensuremath{{\mathrm q}\bar{\mathrm q}{\mathrm g}}}
\def\ggbar{\ensuremath{{\mathrm g}{\mathrm g}}}
\def\nunu{\mbox{\ensuremath{\nu\bar{\nu}}}}

\def\nnbar{\nunu}
\def\tptm{\mbox{\ensuremath{{\tau}^{+}{\tau}^{-}}}}

\def\lplm{\mbox{\ensuremath{{\l}^{+}{\l}^{-}}}}

\def\to{\mbox{\ensuremath{ \rightarrow \ }}}

\newcommand{\ALEPH}{{ALEPH}}

\newcommand{\LEP}{{LEP}}

\newcommand{\CERN}{{CERN}}
\def\capstyl{\small}
\def\NIM#1#2#3{{ Nucl. Instrum. and Methods}
{\bf A#1 }(#2) #3}

%% file: abstract.tex
\begin{abstract}

A search has been performed for the Standard Model Higgs boson
in the data sample collected with the ALEPH detector at LEP,
at centre-of-mass energies up to 209\,GeV.
An excess of $3\,\sigma$ beyond the background expectation
is found, consistent
with the production of the Higgs boson with a mass near 114\,\Gcs.
Much of this excess is seen in the four-jet analyses, where
three high purity events are selected.
\end{abstract}

%% file: authb.tex
\pagestyle{empty}
\newpage
\small
%
%
\newlength{\saveparskip}
\newlength{\savetextheight}
\newlength{\savetopmargin}
\newlength{\savetextwidth}
\newlength{\saveoddsidemargin}
\newlength{\savetopsep}
\setlength{\saveparskip}{\parskip}
\setlength{\savetextheight}{\textheight}
\setlength{\savetopmargin}{\topmargin}
\setlength{\savetextwidth}{\textwidth}
\setlength{\saveoddsidemargin}{\oddsidemargin}
\setlength{\savetopsep}{\topsep}
%
%
\setlength{\parskip}{0.0cm}
\setlength{\textheight}{25.0cm}
\setlength{\topmargin}{-1.5cm}
\setlength{\textwidth}{16 cm}
\setlength{\oddsidemargin}{-0.0cm}
\setlength{\topsep}{1mm}
\pretolerance=10000
\centerline{\large\bf The ALEPH Collaboration}
\footnotesize
\vspace{0.5cm}
{\raggedbottom
\begin{sloppypar}
\samepage\noindent
R.~Barate,
D.~Decamp,
P.~Ghez,
C.~Goy,
S.~Jezequel,
J.-P.~Lees,
F.~Martin,
E.~Merle,
\mbox{M.-N.~Minard},
B.~Pietrzyk
\nopagebreak
\begin{center}
\parbox{15.5cm}{\sl\samepage
Laboratoire de Physique des Particules (LAPP), IN$^{2}$P$^{3}$-CNRS,
F-74019 Annecy-le-Vieux Cedex, France}
\end{center}\end{sloppypar}
\vspace{2mm}
\begin{sloppypar}
\noindent
S.~Bravo,
M.P.~Casado,
M.~Chmeissani,
J.M.~Crespo,
E.~Fernandez,
M.~Fernandez-Bosman,
Ll.~Garrido,$^{15}$
E.~Graug\'{e}s,
J.~Lopez,
M.~Martinez,
G.~Merino,
R.~Miquel,
Ll.M.~Mir,
A.~Pacheco,
D.~Paneque,
H.~Ruiz
\nopagebreak
\begin{center}
\parbox{15.5cm}{\sl\samepage
Institut de F\'{i}sica d'Altes Energies, Universitat Aut\`{o}noma
de Barcelona, E-08193 Bellaterra (Barcelona), Spain$^{7}$}
\end{center}\end{sloppypar}
\vspace{2mm}
\begin{sloppypar}
\noindent
A.~Colaleo,
D.~Creanza,
N.~De~Filippis,
M.~de~Palma,
G.~Iaselli,
G.~Maggi,
M.~Maggi,$^{1}$
S.~Nuzzo,
A.~Ranieri,
G.~Raso,$^{24}$
F.~Ruggieri,
G.~Selvaggi,
L.~Silvestris,
P.~Tempesta,
A.~Tricomi,$^{3}$
G.~Zito
\nopagebreak
\begin{center}
\parbox{15.5cm}{\sl\samepage
Dipartimento di Fisica, INFN Sezione di Bari, I-70126 Bari, Italy}
\end{center}\end{sloppypar}
\vspace{2mm}
\begin{sloppypar}
\noindent
X.~Huang,
J.~Lin,
Q. Ouyang,
T.~Wang,
Y.~Xie,
R.~Xu,
S.~Xue,
J.~Zhang,
L.~Zhang,
W.~Zhao
\nopagebreak
\begin{center}
\parbox{15.5cm}{\sl\samepage
Institute of High Energy Physics, Academia Sinica, Beijing, The People's
Republic of China$^{8}$}
\end{center}\end{sloppypar}
\vspace{2mm}
\begin{sloppypar}
\noindent
D.~Abbaneo,
P.~Azzurri,
T.~Barklow,$^{30}$
G.~Boix,$^{6}$
O.~Buchm\"uller,
M.~Cattaneo,
F.~Cerutti,
B.~Clerbaux,
G.~Dissertori,
H.~Drevermann,
R.W.~Forty,
M.~Frank,
F.~Gianotti,
T.C.~Greening,
J.B.~Hansen,
J.~Harvey,
D.E.~Hutchcroft,
P.~Janot,
B.~Jost,
M.~Kado,
V.~Lemaitre,
P.~Maley,
P.~Mato,
A.~Minten,
A.~Moutoussi,
F.~Ranjard,
L.~Rolandi,
D.~Schlatter,
M.~Schmitt,$^{20}$
O.~Schneider,$^{2}$
P.~Spagnolo,
W.~Tejessy,
F.~Teubert,
E.~Tournefier,$^{26}$
A.~Valassi,
J.J.~Ward,
A.E.~Wright
\nopagebreak
\begin{center}
\parbox{15.5cm}{\sl\samepage
European Laboratory for Particle Physics (CERN), CH-1211 Geneva 23,
Switzerland}
\end{center}\end{sloppypar}
\vspace{2mm}
\begin{sloppypar}
\noindent
Z.~Ajaltouni,
F.~Badaud,
S.~Dessagne,
A.~Falvard,
D.~Fayolle,
P.~Gay,
P.~Henrard,
J.~Jousset,
B.~Michel,
S.~Monteil,
\mbox{J-C.~Montret},
D.~Pallin,
J.M.~Pascolo,
P.~Perret,
F.~Podlyski
\nopagebreak
\begin{center}
\parbox{15.5cm}{\sl\samepage
Laboratoire de Physique Corpusculaire, Universit\'e Blaise Pascal,
IN$^{2}$P$^{3}$-CNRS, Clermont-Ferrand, F-63177 Aubi\`{e}re, France}
\end{center}\end{sloppypar}
\vspace{2mm}
\begin{sloppypar}
\noindent
J.D.~Hansen,
J.R.~Hansen,
P.H.~Hansen,
B.S.~Nilsson,
A.~W\"a\"an\"anen
\nopagebreak
\begin{center}
\parbox{15.5cm}{\sl\samepage
Niels Bohr Institute, 2100 Copenhagen, DK-Denmark$^{9}$}
\end{center}\end{sloppypar}
\vspace{2mm}
\begin{sloppypar}
\noindent
G.~Daskalakis,
A.~Kyriakis,
C.~Markou,
E.~Simopoulou,
A.~Vayaki
\nopagebreak
\begin{center}
\parbox{15.5cm}{\sl\samepage
Nuclear Research Center Demokritos (NRCD), GR-15310 Attiki, Greece}
\end{center}\end{sloppypar}
\vspace{2mm}
\begin{sloppypar}
\noindent
A.~Blondel,$^{12}$
\mbox{J.-C.~Brient},
F.~Machefert,
A.~Roug\'{e},
M.~Swynghedauw,
R.~Tanaka
\linebreak
H.~Videau
\nopagebreak
\begin{center}
\parbox{15.5cm}{\sl\samepage
Laboratoire de Physique Nucl\'eaire et des Hautes Energies, Ecole
Polytechnique, IN$^{2}$P$^{3}$-CNRS, \mbox{F-91128} Palaiseau Cedex, France}
\end{center}\end{sloppypar}
\vspace{2mm}
\begin{sloppypar}
\noindent
E.~Focardi,
G.~Parrini,
K.~Zachariadou
\nopagebreak
\begin{center}
\parbox{15.5cm}{\sl\samepage
Dipartimento di Fisica, Universit\`a di Firenze, INFN Sezione di Firenze,
I-50125 Firenze, Italy}
\end{center}\end{sloppypar}
\vspace{2mm}
\begin{sloppypar}
\noindent
A.~Antonelli,
M.~Antonelli,
G.~Bencivenni,
G.~Bologna,$^{4}$
F.~Bossi,
P.~Campana,
G.~Capon,
V.~Chiarella,
P.~Laurelli,
G.~Mannocchi,$^{5}$
F.~Murtas,
G.P.~Murtas,
L.~Passalacqua,
M.~Pepe-Altarelli$^{25}$
\nopagebreak
\begin{center}
\parbox{15.5cm}{\sl\samepage
Laboratori Nazionali dell'INFN (LNF-INFN), I-00044 Frascati, Italy}
\end{center}\end{sloppypar}
\vspace{2mm}
\begin{sloppypar}
\noindent
M.~Chalmers,
A.W.~Halley,
J.~Kennedy,
J.G.~Lynch,
P.~Negus,
V.~O'Shea,
B.~Raeven,
D.~Smith,
P.~Teixeira-Dias,
A.S.~Thompson
\nopagebreak
\begin{center}
\parbox{15.5cm}{\sl\samepage
Department of Physics and Astronomy, University of Glasgow, Glasgow G12
8QQ,United Kingdom$^{10}$}
\end{center}\end{sloppypar}
\begin{sloppypar}
\noindent
R.~Cavanaugh,
S.~Dhamotharan,
C.~Geweniger,
P.~Hanke,
V.~Hepp,
E.E.~Kluge,
G.~Leibenguth,
A.~Putzer,
K.~Tittel,
S.~Werner,$^{19}$
M.~Wunsch$^{19}$
\nopagebreak
\begin{center}
\parbox{15.5cm}{\sl\samepage
Kirchhoff-Institut f\"ur Physik, Universit\"at Heidelberg, D-69120
Heidelberg, Germany$^{16}$}
\end{center}\end{sloppypar}
\vspace{2mm}
\begin{sloppypar}
\noindent
R.~Beuselinck,
D.M.~Binnie,
W.~Cameron,
G.~Davies,
P.J.~Dornan,
M.~Girone,$^{1}$
N.~Marinelli,
J.~Nowell,
H.~Przysiezniak,
J.K.~Sedgbeer,
J.C.~Thompson,$^{14}$
E.~Thomson,$^{23}$
R.~White
\nopagebreak
\begin{center}
\parbox{15.5cm}{\sl\samepage
Department of Physics, Imperial College, London SW7 2BZ,
United Kingdom$^{10}$}
\end{center}\end{sloppypar}
\vspace{2mm}
\begin{sloppypar}
\noindent
V.M.~Ghete,
P.~Girtler,
E.~Kneringer,
D.~Kuhn,
G.~Rudolph
\nopagebreak
\begin{center}
\parbox{15.5cm}{\sl\samepage
Institut f\"ur Experimentalphysik, Universit\"at Innsbruck, A-6020
Innsbruck, Austria$^{18}$}
\end{center}\end{sloppypar}
\vspace{2mm}
\begin{sloppypar}
\noindent
E.~Bouhova-Thacker,
C.K.~Bowdery,
D.P.~Clarke,
G.~Ellis,
A.J.~Finch,
F.~Foster,
G.~Hughes,
R.W.L.~Jones,$^{1}$
M.R.~Pearson,
N.A.~Robertson,
M.~Smizanska
\nopagebreak
\begin{center}
\parbox{15.5cm}{\sl\samepage
Department of Physics, University of Lancaster, Lancaster LA1 4YB,
United Kingdom$^{10}$}
\end{center}\end{sloppypar}
\vspace{2mm}
\begin{sloppypar}
\noindent
I.~Giehl,
F.~H\"olldorfer,
K.~Jakobs,
K.~Kleinknecht,
M.~Kr\"ocker,
A.-S.~M\"uller,
H.-A.~N\"urnberger,
G.~Quast,$^{1}$
B.~Renk,
E.~Rohne,
H.-G.~Sander,
S.~Schmeling,
H.~Wachsmuth,
C.~Zeitnitz,
T.~Ziegler
\nopagebreak
\begin{center}
\parbox{15.5cm}{\sl\samepage
Institut f\"ur Physik, Universit\"at Mainz, D-55099 Mainz, Germany$^{16}$}
\end{center}\end{sloppypar}
\vspace{2mm}
\begin{sloppypar}
\noindent
A.~Bonissent,
J.~Carr,
P.~Coyle,
C.~Curtil,
A.~Ealet,
D.~Fouchez,
O.~Leroy,
T.~Kachelhoffer,
P.~Payre,
D.~Rousseau,
A.~Tilquin
\nopagebreak
\begin{center}
\parbox{15.5cm}{\sl\samepage
Centre de Physique des Particules de Marseille, Univ M\'editerran\'ee,
IN$^{2}$P$^{3}$-CNRS, F-13288 Marseille, France}
\end{center}\end{sloppypar}
\vspace{2mm}
\begin{sloppypar}
\noindent
M.~Aleppo,
S.~Gilardoni,
F.~Ragusa
\nopagebreak
\begin{center}
\parbox{15.5cm}{\sl\samepage
Dipartimento di Fisica, Universit\`a di Milano e INFN Sezione di
Milano, I-20133 Milano, Italy.}
\end{center}\end{sloppypar}
\vspace{2mm}
\begin{sloppypar}
\noindent
A.~David,
H.~Dietl,
G.~Ganis,$^{27}$
A.~Heister,
K.~H\"uttmann,
G.~L\"utjens,
C.~Mannert,
W.~M\"anner,
\mbox{H.-G.~Moser},
S.~Schael,
R.~Settles,$^{1}$
H.~Stenzel,
G.~Wolf
\nopagebreak
\begin{center}
\parbox{15.5cm}{\sl\samepage
Max-Planck-Institut f\"ur Physik, Werner-Heisenberg-Institut,
D-80805 M\"unchen, Germany\footnotemark[16]}
\end{center}\end{sloppypar}
\vspace{2mm}
\begin{sloppypar}
\noindent
J.~Boucrot,$^{1}$
O.~Callot,
M.~Davier,
L.~Duflot,
\mbox{J.-F.~Grivaz},
Ph.~Heusse,
A.~Jacholkowska,$^{1}$
L.~Serin,
\mbox{J.-J.~Veillet},
I.~Videau,
J.-B.~de~Vivie~de~R\'egie,$^{28}$
C.~Yuan,
D.~Zerwas
\nopagebreak
\begin{center}
\parbox{15.5cm}{\sl\samepage
Laboratoire de l'Acc\'el\'erateur Lin\'eaire, Universit\'e de Paris-Sud,
IN$^{2}$P$^{3}$-CNRS, F-91898 Orsay Cedex, France}
\end{center}\end{sloppypar}
\vspace{2mm}
\begin{sloppypar}
\noindent
G.~Bagliesi,
T.~Boccali,
G.~Calderini,
V.~Ciulli,
L.~Fo\`a,
A.~Giammanco,
A.~Giassi,
F.~Ligabue,
A.~Messineo,
F.~Palla,$^{1}$
G.~Rizzo,
G.~Sanguinetti,
A.~Sciab\`a,
G.~Sguazzoni,
R.~Tenchini,$^{1}$
A.~Venturi,
P.G.~Verdini
\samepage
\begin{center}
\parbox{15.5cm}{\sl\samepage
Dipartimento di Fisica dell'Universit\`a, INFN Sezione di Pisa,
e Scuola Normale Superiore, I-56010 Pisa, Italy}
\end{center}\end{sloppypar}
\vspace{2mm}
\begin{sloppypar}
\noindent
G.A.~Blair,
J.~Coles,
G.~Cowan,
M.G.~Green,
L.T.~Jones,
T.~Medcalf,
J.A.~Strong
\nopagebreak
\begin{center}
\parbox{15.5cm}{\sl\samepage
Department of Physics, Royal Holloway \& Bedford New College,
University of London, Surrey TW20 OEX, United Kingdom$^{10}$}
\end{center}\end{sloppypar}
\vspace{2mm}
\begin{sloppypar}
\noindent
R.W.~Clifft,
T.R.~Edgecock,
P.R.~Norton,
I.R.~Tomalin
\nopagebreak
\begin{center}
\parbox{15.5cm}{\sl\samepage
Particle Physics Dept., Rutherford Appleton Laboratory,
Chilton, Didcot, Oxon OX11 OQX, United Kingdom$^{10}$}
\end{center}\end{sloppypar}
\vspace{2mm}
\begin{sloppypar}
\noindent
\mbox{B.~Bloch-Devaux},$^{1}$
D.~Boumediene,
P.~Colas,
B.~Fabbro,
E.~Lan\c{c}on,
\mbox{M.-C.~Lemaire},
E.~Locci,
P.~Perez,
J.~Rander,
\mbox{J.-F.~Renardy},
A.~Rosowsky,
P.~Seager,$^{13}$
A.~Trabelsi,$^{21}$
B.~Tuchming,
B.~Vallage
\nopagebreak
\begin{center}
\parbox{15.5cm}{\sl\samepage
CEA, DAPNIA/Service de Physique des Particules,
CE-Saclay, F-91191 Gif-sur-Yvette Cedex, France$^{17}$}
\end{center}\end{sloppypar}
\vspace{2mm}
\begin{sloppypar}
\noindent
N.~Konstantinidis,
C.~Loomis,
A.M.~Litke,
G.~Taylor
\nopagebreak
\begin{center}
\parbox{15.5cm}{\sl\samepage
Institute for Particle Physics, University of California at
Santa Cruz, Santa Cruz, CA 95064, USA$^{22}$}
\end{center}\end{sloppypar}
\vspace{2mm}
\begin{sloppypar}
\noindent
C.N.~Booth,
S.~Cartwright,
F.~Combley,
P.N.~Hodgson,
M.~Lehto,
L.F.~Thompson
\nopagebreak
\begin{center}
\parbox{15.5cm}{\sl\samepage
Department of Physics, University of Sheffield, Sheffield S3 7RH,
United Kingdom$^{10}$}
\end{center}\end{sloppypar}
\vspace{2mm}
\begin{sloppypar}
\noindent
K.~Affholderbach,
A.~B\"ohrer,
S.~Brandt,
C.~Grupen,
J.~Hess,
A.~Misiejuk,
G.~Prange,
U.~Sieler
\nopagebreak
\begin{center}
\parbox{15.5cm}{\sl\samepage
Fachbereich Physik, Universit\"at Siegen, D-57068 Siegen, Germany$^{16}$}
\end{center}\end{sloppypar}
\vspace{2mm}
\begin{sloppypar}
\noindent
C.~Borean,
G.~Giannini,
B.~Gobbo
\nopagebreak
\begin{center}
\parbox{15.5cm}{\sl\samepage
Dipartimento di Fisica, Universit\`a di Trieste e INFN Sezione di Trieste,
I-34127 Trieste, Italy}
\end{center}\end{sloppypar}
\vspace{2mm}
\begin{sloppypar}
\noindent
H.~He,
J.~Putz,
J.~Rothberg,
S.~Wasserbaech
\nopagebreak
\begin{center}
\parbox{15.5cm}{\sl\samepage
Experimental Elementary Particle Physics, University of Washington, Seattle,
WA 98195 U.S.A.}
\end{center}\end{sloppypar}
\vspace{2mm}
\begin{sloppypar}
\noindent
S.R.~Armstrong,
K.~Cranmer,
P.~Elmer,
D.P.S.~Ferguson,
Y.~Gao,$^{29}$
S.~Gonz\'{a}lez,
O.J.~Hayes,
H.~Hu,
S.~Jin,
J.~Kile,
P.A.~McNamara III,
J.~Nielsen,
W.~Orejudos,
Y.B.~Pan,
Y.~Saadi,
I.J.~Scott,
N.~Shao,
\mbox{J.H.~von~Wimmersperg-Toeller}, 
J.~Walsh,
W.~Wiedenmann,
J.~Wu,
Sau~Lan~Wu,
X.~Wu,
G.~Zobernig
\nopagebreak
\begin{center}
\parbox{15.5cm}{\sl\samepage
Department of Physics, University of Wisconsin, Madison, WI 53706,
USA$^{11}$}
\end{center}\end{sloppypar}
}
\footnotetext[1]{Also at CERN, 1211 Geneva 23, Switzerland.}
\footnotetext[2]{Now at Universit\'e de Lausanne, 1015 Lausanne, Switzerland.}
\footnotetext[3]{Also at Dipartimento di Fisica di Catania and INFN Sezione di
 Catania, 95129 Catania, Italy.}
\footnotetext[4]{Deceased.}
\footnotetext[5]{Also Istituto di Cosmo-Geofisica del C.N.R., Torino,
Italy.}
\footnotetext[6]{Supported by the Commission of the European Communities,
contract ERBFMBICT982894.}
\footnotetext[7]{Supported by CICYT, Spain.}
\footnotetext[8]{Supported by the National Science Foundation of China.}
\footnotetext[9]{Supported by the Danish Natural Science Research Council.}
\footnotetext[10]{Supported by the UK Particle Physics and Astronomy Research
Council.}
\footnotetext[11]{Supported by the US Department of Energy, grant
DE-FG0295-ER40896.}
\footnotetext[12]{Now at Departement de Physique Corpusculaire, Universit\'e de
Gen\`eve, 1211 Gen\`eve 4, Switzerland.}
\footnotetext[13]{Supported by the Commission of the European Communities,
contract ERBFMBICT982874.}
\footnotetext[14]{Also at Rutherford Appleton Laboratory, Chilton, Didcot, UK.}
\footnotetext[15]{Permanent address: Universitat de Barcelona, 08208 Barcelona,
Spain.}
\footnotetext[16]{Supported by the Bundesministerium f\"ur Bildung,
Wissenschaft, Forschung und Technologie, Germany.}
\footnotetext[17]{Supported by the Direction des Sciences de la
Mati\`ere, C.E.A.}
\footnotetext[18]{Supported by the Austrian Ministry for Science and Transport.}
\footnotetext[19]{Now at SAP AG, 69185 Walldorf, Germany}
\footnotetext[20]{Now at Harvard University, Cambridge, MA 02138, U.S.A.}
\footnotetext[21]{Now at D\'epartement de Physique, Facult\'e des Sciences de Tunis, 1060 Le Belv\'ed\`ere, Tunisia.}
\footnotetext[22]{Supported by the US Department of Energy,
grant DE-FG03-92ER40689.}
\footnotetext[23]{Now at Department of Physics, Ohio State University, Columbus, OH 43210-1106, U.S.A.}
\footnotetext[24]{Also at Dipartimento di Fisica e Tecnologie Relative, Universit\`a di Palermo, Palermo, Italy.}
\footnotetext[25]{Now at CERN, 1211 Geneva 23, Switzerland.}
\footnotetext[26]{Now at ISN, Institut des Sciences Nucl\'eaires, 53 Av. des Martyrs, 38026 Grenoble, France.}
\footnotetext[27]{Now at Universit\`a degli Studi di Roma Tor Vergata, Dipartimento di Fisica, 00133 Roma, Italy.}
\footnotetext[28]{Now at Centre de Physique des Particules de Marseille,Univ M\'editerran\'ee, F-13288 Marseille, France.}
\footnotetext[29]{Also at Department of Physics, Tsinghua University, Beijing, The People's Republic of China.}
\footnotetext[30]{Also at SLAC, Stanford, CA 94309, U.S.A.}
%
\setlength{\parskip}{\saveparskip}
\setlength{\textheight}{\savetextheight}
\setlength{\topmargin}{\savetopmargin}
\setlength{\textwidth}{\savetextwidth}
\setlength{\oddsidemargin}{\saveoddsidemargin}
\setlength{\topsep}{\savetopsep}
\normalsize
\newpage
\pagestyle{plain}
\setcounter{page}{1}

%% file: introduction.tex
\section{Introduction}

This letter presents results on the search for the Standard Model
Higgs boson~\cite{higgs} using the data collected by ALEPH at LEP in
the year 2000.  Similar analyses on previous years' data, with
centre-of-mass energies up to 202\,GeV, have shown no evidence for a
signal~\cite{Paper1999}.  A lower mass limit of 107.7\,\Gcs\ was set
at the 95\% confidence level.

\begin{figure}[hbtp]
\begin{center}
\begin{fmffile}{feyn_higgs}
\begin{fmfgraph*}(60,35)
\fmfleft{i1,i2} \fmfright{o1,o2}
\fmf{fermion, label=$e^-$}{i1,v1}
\fmf{fermion, label=$e^+$}{v1,i2} 
\fmf{boson,label= $Z^\star$}{v1,v2} \fmfdot{v1,v2}
\fmf{dashes, label= $H$}{v2,o2}
\fmf{boson, label= $Z$}{v2,o1}
\end{fmfgraph*}
\hspace{1cm}
\begin{fmfgraph*}(60,30)
\fmfleft{i1,i2} \fmfright{o1,o2,o3}
\fmfforce{.5w,0}{v1}
\fmfforce{.5w,h}{v3}
\fmf{fermion, label=$e^-$}{i1,v1}
\fmf{fermion, label=$\nu_e,,e^-$}{v1,o1}
\fmf{fermion, label=$e^+$}{v3,i2}
\fmf{fermion, label=$\bar{\nu_e},,e^+$}{o3,v3} 
\fmf{boson,label=$W^-,,Z$}{v1,v2} \fmfdot{v1,v2,v3}
\fmf{boson,label=$W^+,,Z$}{v2,v3}
\fmf{dashes, label= $H$}{v2,o2}
\end{fmfgraph*}
\end{fmffile}
\end{center}
\begin{flushleft}
{\vspace{-2.1cm} {\Large  \hspace{1.cm} {(a)}
\hspace{6.5cm} {(b)}} \vspace{2.1cm}}
\end{flushleft}
\centering
\vspace{-1.0cm}
\caption
{\capstyl Feynman diagrams of Higgs boson production at LEP through the
(a) Higgsstrahlung and (b) gauge boson fusion processes.}
\label{feynman}
\end{figure}
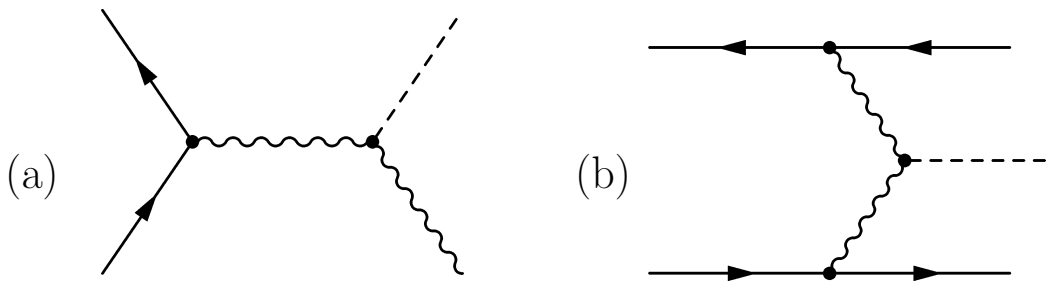

The ALEPH detector, described fully in Ref.~\cite{aleph},
is a general purpose detector composed of tracking,
vertexing, and calorimetry subdetectors.
This search looks for a Higgs boson produced in
association with a Z boson through the Higgsstrahlung process,
$\epem\,\to\H\Z$~\cite{ellis}. This process is supplemented
by a small contribution from the W and Z vector boson fusion processes,
which produce a Higgs boson and either a pair of neutrinos or electrons
in the final state~\cite{fusion}.
The Feynman diagrams for these processes are shown
in Fig.~\ref{feynman}.

\begin{figure}[hbtp]
\begin{center}
\begin{picture}(170,80)
\put(40,0){\epsfxsize=82mm\epsfbox{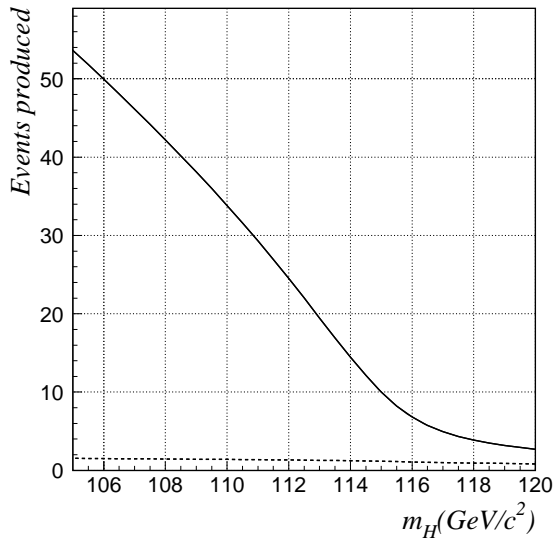}}
\end{picture}
\vspace{-1.0cm}
\caption{\capstyl
The expected number of Standard Model Higgs boson events produced in
the year 2000 data
as a function of the Higgs boson mass (solid curve).
The dashed curve shows the contribution of the
boson fusion processes, including their interference with the
Higgsstrahlung process.
\label{xsect}} 
\end{center}
\end{figure}

The centre-of-mass energies for the
216.1\,\invpb\ of data collected in the year 2000 range from 200\,GeV to
209\,GeV, with the majority of the data collected around
205.1\,GeV (72\,\invpb) and 206.7\,GeV (107\,\invpb).
Figure~\ref{xsect} shows the number of Higgs boson events expected to be
produced in this sample as a function of the Higgs boson mass. For a
mass of 114\,\Gcs, 14.4 signal events are expected to be produced.
The Higgs boson at this mass
predominantly decays into \bbbar\ quark pairs (74\%) and tau lepton
pairs (7\%). The overall selection efficiency is typically 50\%.

The purpose of this letter is to report the observation of an excess
which is consistent with the production of the Standard Model
Higgs boson with a mass near 114\,\Gcs. 
These results are based upon events reconstructed using preliminary
detector calibrations.
The results obtained after the final event processing, a possible
reoptimization of some analyses, and a more detailed investigation of
systematic effects, will be reported in a forthcoming publication. It
has been verified that the significance of the most signal-like events
is not affected by the final processing.


%% file: analyses.tex
\section{Event Selection}

The analyses designed to search for the Standard Model Higgs boson
address most of the final states arising from the reaction
\epem\,\to\h\Z: the four-jet
final state (\h\qqbar), the missing energy final state (\h\nnbar), the
lepton pair final state (\h\lplm\ where $\ell$ denotes an electron or muon),
and the tau lepton final state (\h\tptm\ and \h\,\to\tptm, \Z\,\to\qqbar).
As in Ref.~\cite{Paper1999}, the Higgs boson search was conducted using both a
neural-network-based stream (denoted ``NN'') and a cut-based stream
(``cut''). Alternative analyses are used in the searches for four-jet and for 
missing energy final states, while the searches are identical in
both streams for the lepton pair and tau lepton final states. 
All of the analysis selection
criteria for the results presented here
were fixed before the data taking period began.

These analyses follow closely those designed for the data collected
in 1999~\cite{Paper1999}. In particular, the b tagging neural network 
described in Ref.~\cite{btag}
is used, based upon the b-hadron lifetime, mass and
semi-leptonic decays. Its output ranges from 0 to 1,
where a value of 1 indicates a well b tagged jet. The
four-jet and tau lepton final state analyses are identical to
those of 1999,
while the missing energy and
lepton pair final state analyses have had the following improvements:

\begin{itemize}
\item The two neural network analyses used in the search for the missing
energy final state in the NN stream
have been replaced by a single neural network with three output 
classes.
This neural network is trained to identify three types of events: the signal,
the \qqbar\ background, and the \WW\ background. With this technique
the benefits of the two previous approaches are merged 
without any loss in performance.

\item The missing energy analysis for the cut stream has an
improved rejection of three-jet events from the \qqg$(\gamma)$ and
$\qq\gamma(\gamma)$ processes. A jet algorithm~\cite{durham} is applied to 
form three jets. Events from \qqg$(\gamma)$ are removed by cuts on the minimum 
angle and on the minimum distance~\cite{durham} between any two jets.
Three-jet events originating from the $\qq\gamma(\gamma)$ process with a
photon in the detector are removed if any of the three jets is 
predominantly electromagnetic in origin.

\item The analysis for the lepton pair final state has three improvements:
1) an increased efficiency
in the identification of events in which the Higgs boson decays 
to $\tau$ leptons,
2) reduced expected background for events
with isolated photons, and 3) an increased sensitivity above the nominal
kinematic limit, achieved by relaxing requirements on the measured 
Z boson mass.

\end{itemize}

The four-jet NN and cut analyses differ in their
method to choose the best jet pairing.
The four-jet cut analysis chooses the pairing, as described in
Ref.~\cite{Paper1998}, based upon the decay angles
of the Z and Higgs bosons.
The NN analysis includes these two variables in the neural network
and selects the jet pairing with the largest neural network output
value, thereby effectively also using the b tagging and
Z boson mass information.

In the four-jet analysis, a 4C-kinematic fit is 
performed, in which energy and momentum conservation are imposed. 
The reconstructed Higgs boson mass $m_{REC}$ is calculated as 
$m_{12}+m_{34}-m_{\mathrm Z}$, where $m_{12}$ and $m_{34}$ are the fitted Z 
and Higgs boson masses.
In the missing energy final state, the Higgs boson mass is 
reconstructed by a rescaling of the hadronic jets such that the 
missing mass is the Z boson mass.
In the lepton pair final state, it is calculated as the 
mass recoiling against the pair of leptons.
In the tau lepton final state, it results 
from a kinematic fit, with the Z mass constraint imposed either on the tau 
pair or on the hadronic system.

Fully simulated samples of signal and background processes were
produced with the same generators used in Ref.~\cite{Paper1999}. The sizes
of the simulated samples correspond to at least 50 times the collected
luminosity.
For each analysis, the expected numbers of signal and background events
and the number of observed candidates are given in Table~\ref{bigtable}.

\begin{table}[t]
\begin{center}
\begin{tabular}{|c|c|r@{$\pm$}l|r@{$\pm$}l|r@{$\pm$}l|r@{$\pm$}l|c|c|}
\hline\hline
Analysis & Signal &
\multicolumn{8}{c|}{Background Events} & Events & Expected \\
 & Events &
\multicolumn{8}{c|}{Expected} & Obs. & Significance \\

\cline{3-10}
 & Expected  & \multicolumn{2}{c|}{\Z\Z} &
\multicolumn{2}{c|}{\W\W} & \multicolumn{2}{c|}{\rule{0cm}{0.45cm}\ffbar} &
\multicolumn{2}{c|}{Total} &  & ($\sigma$) \\
\hline
\hline
\h\qqbar\ (NN) &
4.5  &                     
23.0  & 1.0 &                     
8.6  & 0.6 &                     
15.3  & 1.7 &                     
46.9 & 2.1 &                     
52   &                           
1.6 \\                          
\hline
\h\qqbar\ (Cut) &
2.9  &                     
12.6 & 0.7 &                     
3.2  & 0.2 &                     
7.9  & 0.7 &                     
23.7 & 1.0 &                     
31   &                           
1.3 \\                          
\hline
\h\nnbar\ (NN) &
1.4  &                     
13.5 & 0.7 &                     
22.0 & 1.1 &                     
2.0 & 0.4 &                     
37.5 & 1.4 &                     
38   &                           
0.8 \\                          
\hline
\h\nnbar\ (Cut) &
1.3  &                     
9.9 & 1.1 &                     
8.8 & 1.7 &                     
1.0 & 0.3 &                     
19.7 & 2.0 &                     
20   &                           
0.7 \\                          
\hline
\h\lplm &
0.7  &                     
26.4 & 0.3 &                     
2.4 & 0.1 &                     
1.8 & 0.3 &                     
30.6 & 0.4 &                     
29   &                           
0.8 \\                          
\hline
\tptm\qqbar &
0.4  &                     
6.4 & 0.3 &                     
6.2 & 0.3 &                     
1.0 & 0.3 &                     
13.6 & 0.5 &                     
15   &                           
0.4 \\                          
\hline
\hline
NN Total &
7.0  &                     
69.3 & 1.3 &                     
39.2 & 1.3 &                     
20.1 & 1.8 &                     
128.7 & 2.6 &                     
134   &                           
2.1 \\                          
\hline
Cut Total &
5.3  &                     
55.3 & 1.4 &                     
20.6 & 1.7 &                     
11.7 & 0.9 &                     
87.6 & 2.4 &                     
95  &                           
1.8 \\                          
\hline
\hline
\end{tabular}
\caption{\capstyl The number of signal and background events expected, and 
the number of candidate events observed in the year 2000 data. For each channel
the systematic error on the background is indicated.
The expected background is divided into
\Z\Z\ (including $\Z\epem$ and \Z\nunu), \W\W\ (including $\W\e\nu$), 
and \ffbar\ (including $\gamma\gamma~\to\ffbar$).
The expectation for the signal and its significance (Section~3)
are computed for
a Higgs boson with a mass of 114\,\Gcs.
The numbers from the \h\lplm\ and \tptm\qqbar\ analyses are included
in both the NN and cut totals.
\label{bigtable}}
\end{center}
\end{table}

\begin{figure}
\begin{center}
\begin{picture}(170,80)
\put(60,65){\Large \bf (a)}
\put(12,75){\Large ALEPH}
\put(0,0){\epsfxsize=82mm\epsfbox{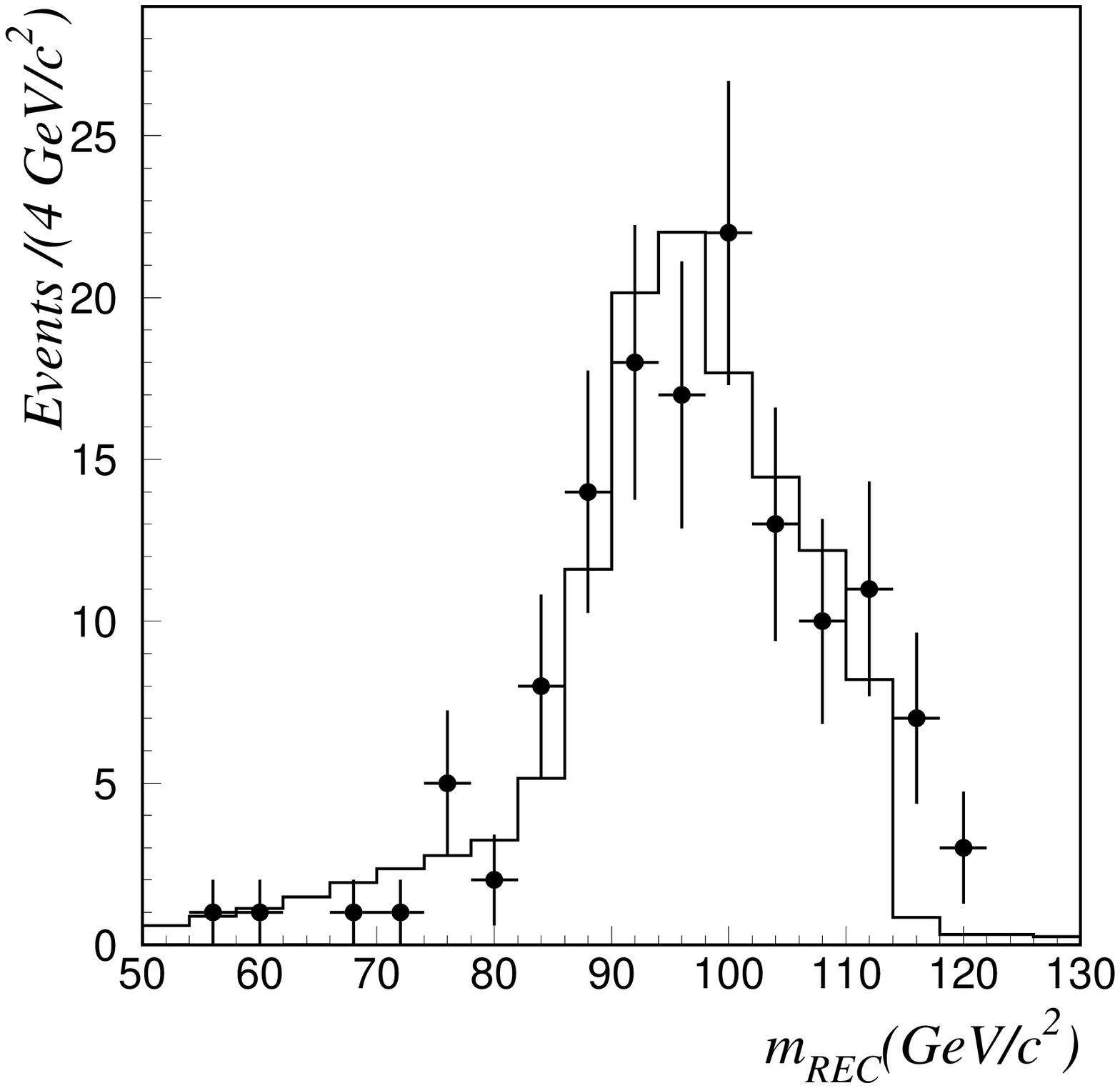}}
\put(145,65){\Large \bf (b)}
\put(97,75){\Large ALEPH}
\put(85,0){\epsfxsize=82mm\epsfbox{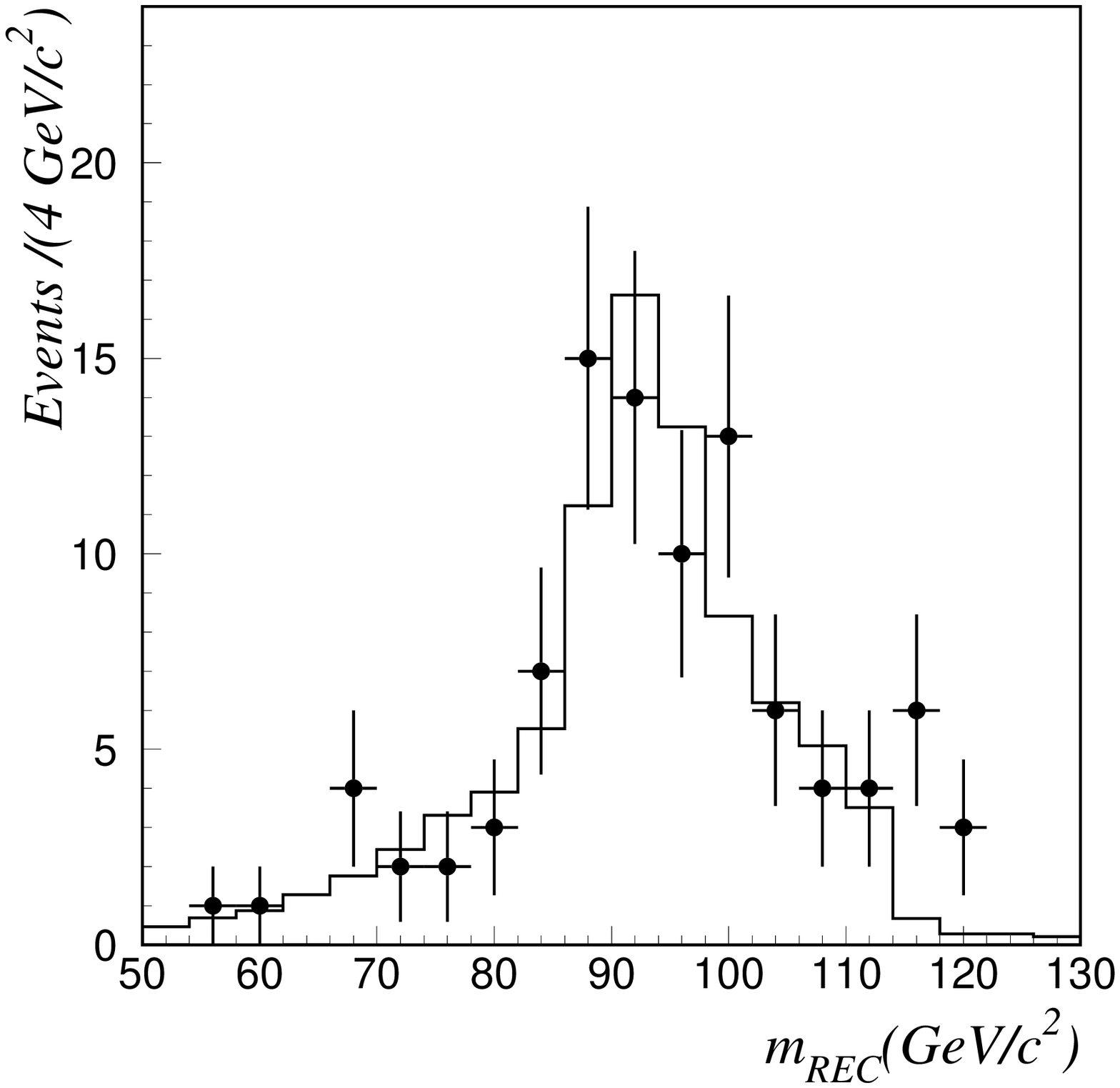}}
\end{picture}
\caption{\capstyl Distributions of the reconstructed Higgs boson mass for the  
data collected in 2000 (dots with error bars) and the expected background 
(histogram) for the (a) NN and (b) cut streams.
\label{MassPlots}} 
\end{center}
\end{figure}

For the NN (cut) searches, a total of 134 (95)
events are selected in the data,
while 128.7 (87.6) events
are expected from Standard Model background processes.
Figures~\ref{MassPlots}a (NN) and~\ref{MassPlots}b (cut) show the
distributions of the reconstructed Higgs boson mass for the data and
the expected background. Although there is good agreement
between the data (dots) and the expected background (histogram) in
the total number of events, an excess of data events can be seen in both the
NN and cut distributions for large reconstructed
Higgs boson masses.

%% file: cl_results2.tex
\section{Confidence Level Results}

The mass is not the only information which allows Higgs boson production
to be distinguished from background.  Additional information is taken
into account in the likelihood ratio $Q = L_{s+b}/L_{b}$,
where $L_{b}$ is the likelihood of the background hypothesis,
and $L_{s+b}$ is the likelihood when a specific Higgs boson signal
is added to the background.
The likelihood ratio measures the compatibility
of the experiment with a particular signal mass hypothesis:

\[
Q = \frac{L_{s+b}}{L_{b}} = \frac{e^{-(s+b)}}{e^{-b}}
\prod_{i=1}^{n_{obs}} \frac{s f_{s}(\vec{X}_{i}) + b f_{b}(\vec{X}_{i})}
{b f_b(\vec{X}_{i})}
\]
where $s$ and $b$ are the total numbers of signal and background events
expected. Neglecting the $f_s$ and $f_b$ terms, this is simply the ratio of
the Poisson probabilities to observe $n_{obs}$ events for the
signal-plus-background and background-only hypotheses.
The functions $f_s$ and $f_b$ are the probability densities that a signal or
background event will be found in a given final state with the set of
values $\vec{X}_i$ which includes the reconstructed mass and possibly
a second discriminant.

The four-jet NN analysis uses the
neural network output as a second discriminant, while the missing energy NN
and lepton pair selections
use the sum of the b~tagging neural network output values of the
hadronic jets as a second discriminant.
The four-jet cut, missing energy cut, and tau lepton analyses use only
the reconstructed Higgs boson mass as a discriminant.

A small correlation between the
neural network output and the reconstructed Higgs boson mass for both
the signal and the background distributions was observed in the four-jet
NN analysis. The effect of this correlation has been taken into
account.  No correlation was found between the
b tagging distributions and the reconstructed Higgs boson mass in 
the missing energy NN and lepton pair analyses.

The compatibility of an experiment with a given hypothesis is determined
from the expected distribution of the likelihood ratio by calculating
the probability of obtaining a likelihood ratio smaller than the one
observed.  This probability, called the confidence level (CL), depends
upon the hypothesized Higgs boson mass for both the signal-plus-background
and the background-only hypotheses.
If the hypothesis being tested is true, then the distribution of
possible confidence levels is equally distributed between 0 and 1,
with a median value of 0.5.
A signal is expected to produce an excess relative to the expected
background, which would appear as a dip in $1 - c_b$, where $c_b$ is
the confidence level for the background-only hypothesis.

\begin{figure}
\begin{center}
\begin{picture}(170,80)
\put(60,65){\Large \bf (a)}
\put(12,75){\Large ALEPH}
\put(0,0){\epsfxsize=82mm\epsfbox{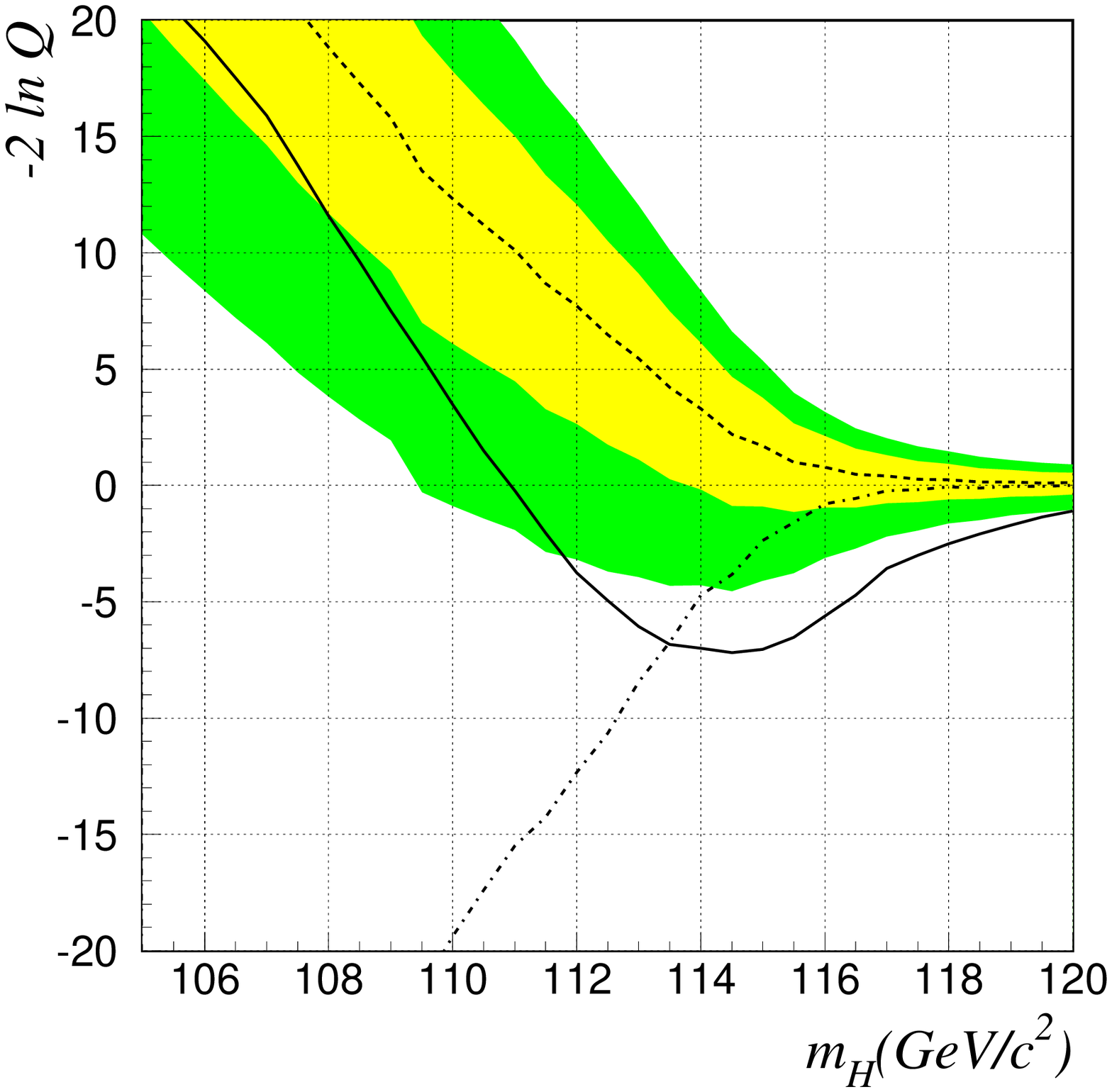}}
\put(145,65){\Large \bf (b)}
\put(97,75){\Large ALEPH}
\put(85,0){\epsfxsize=82mm\epsfbox{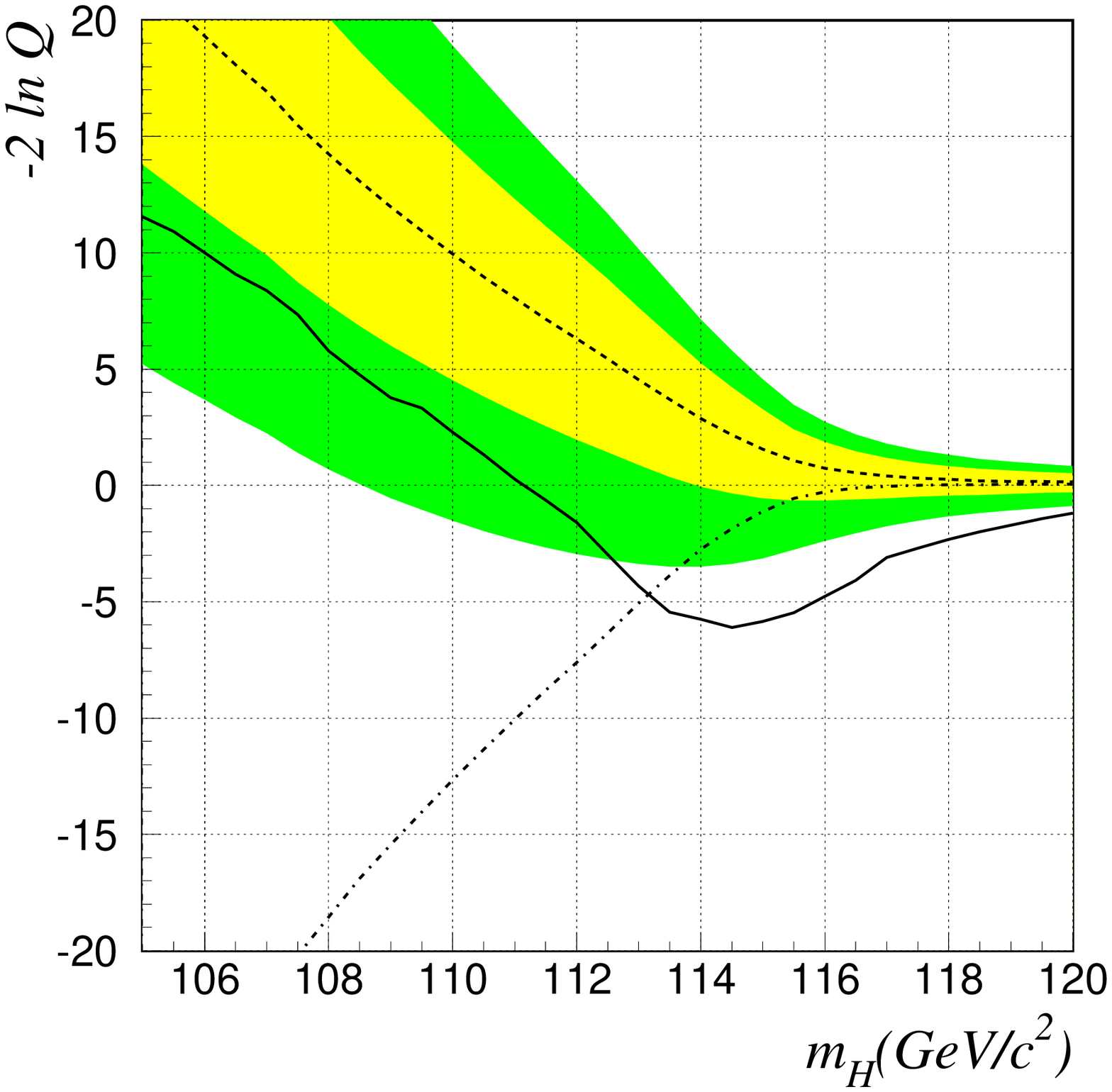}}
\end{picture}
\caption{\capstyl The log-likelihood estimator $-2 \ln{Q}$
for the (a) NN and (b) cut streams as a
function of the mass of the Higgs boson for the
observation (solid) and background-only expectation (dashed).
The light and dark grey regions around the background expectation represent
the one and two sigma bands, respectively.
The dash-dotted
curves show the medians of the log-likelihood
estimator as a function of the Higgs boson mass for the signal hypothesis.
\label{LR}} 
\end{center}
\end{figure}

The 176\,\invpb\ and 237\,\invpb\ of data collected in 1998 and 1999
respectively, with
centre-of-mass energies ranging from 188.6\,GeV to 201.6\,GeV, were
combined with the data collected in the year 2000 to determine
the compatibility of the results with either the background-only or
signal-plus-background hypotheses.
The observed distribution of $-2 \ln{Q}$ is shown as a function of the hypothesized
Higgs boson mass in Fig.~\ref{LR}a (NN) and Fig.~\ref{LR}b (cut).
The likelihood ratio is traditionally shown in the form
$-2 \ln{Q}$ (the log-likelihood estimator) because
of the relationship between the likelihood ratio and chi-squared
distributions, and because when the logarithm is taken, individual
events contribute as a sum of event weights,
$\ln{(1+\frac{s f_{s}}{b f_{b}})}$, which can be examined
individually.  The most likely Higgs boson mass corresponds to the
minimum, observed near 114\,\Gcs.

\begin{figure}
\begin{center}
\begin{picture}(170,80)
\put(15,25){\Large \bf (a)}
\put(12,75){\Large ALEPH}
\put(0,0){\epsfxsize=82mm\epsfbox{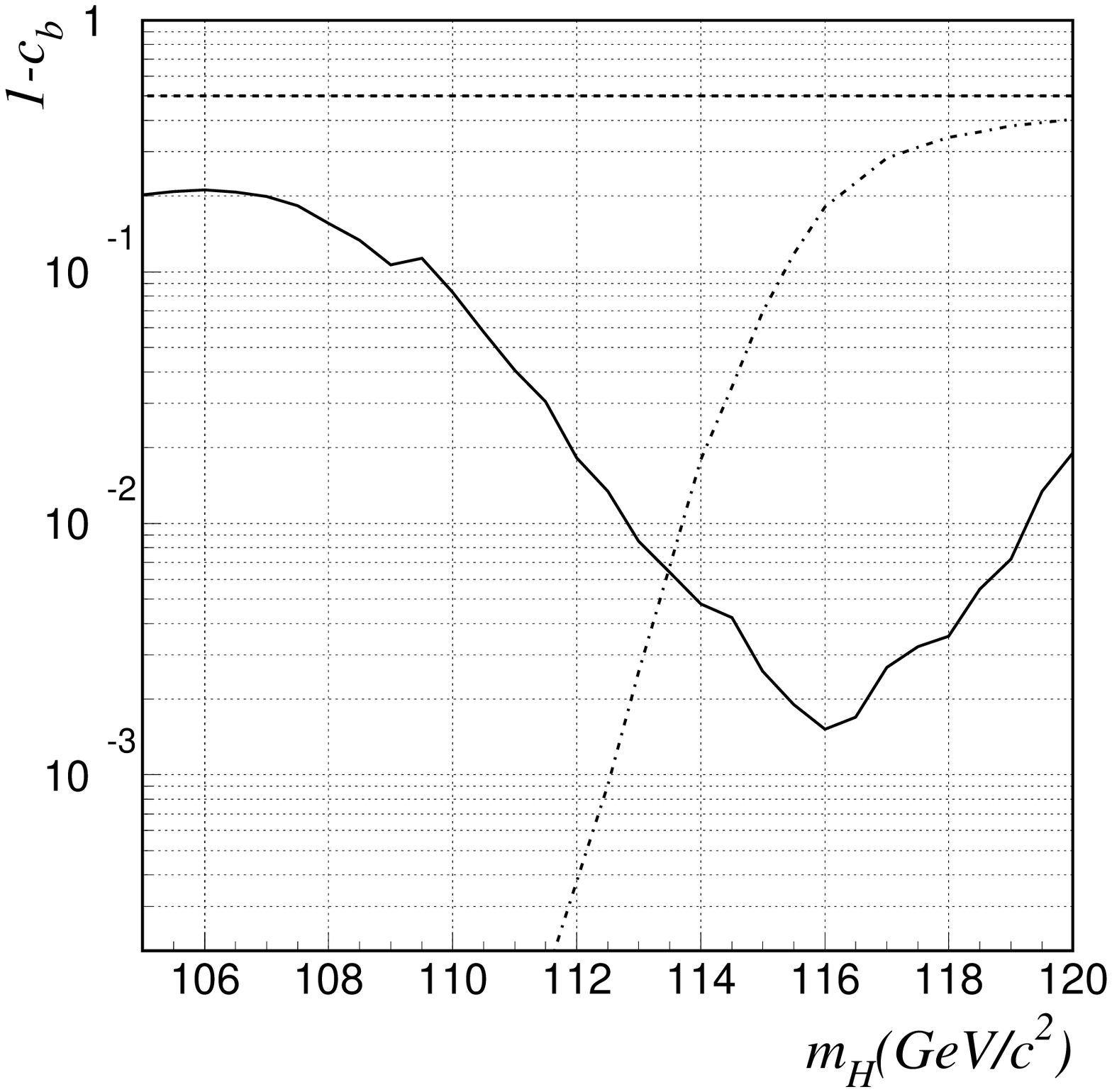}}
\put(100,25){\Large \bf (b)}
\put(97,75){\Large ALEPH}
\put(85,0){\epsfxsize=82mm\epsfbox{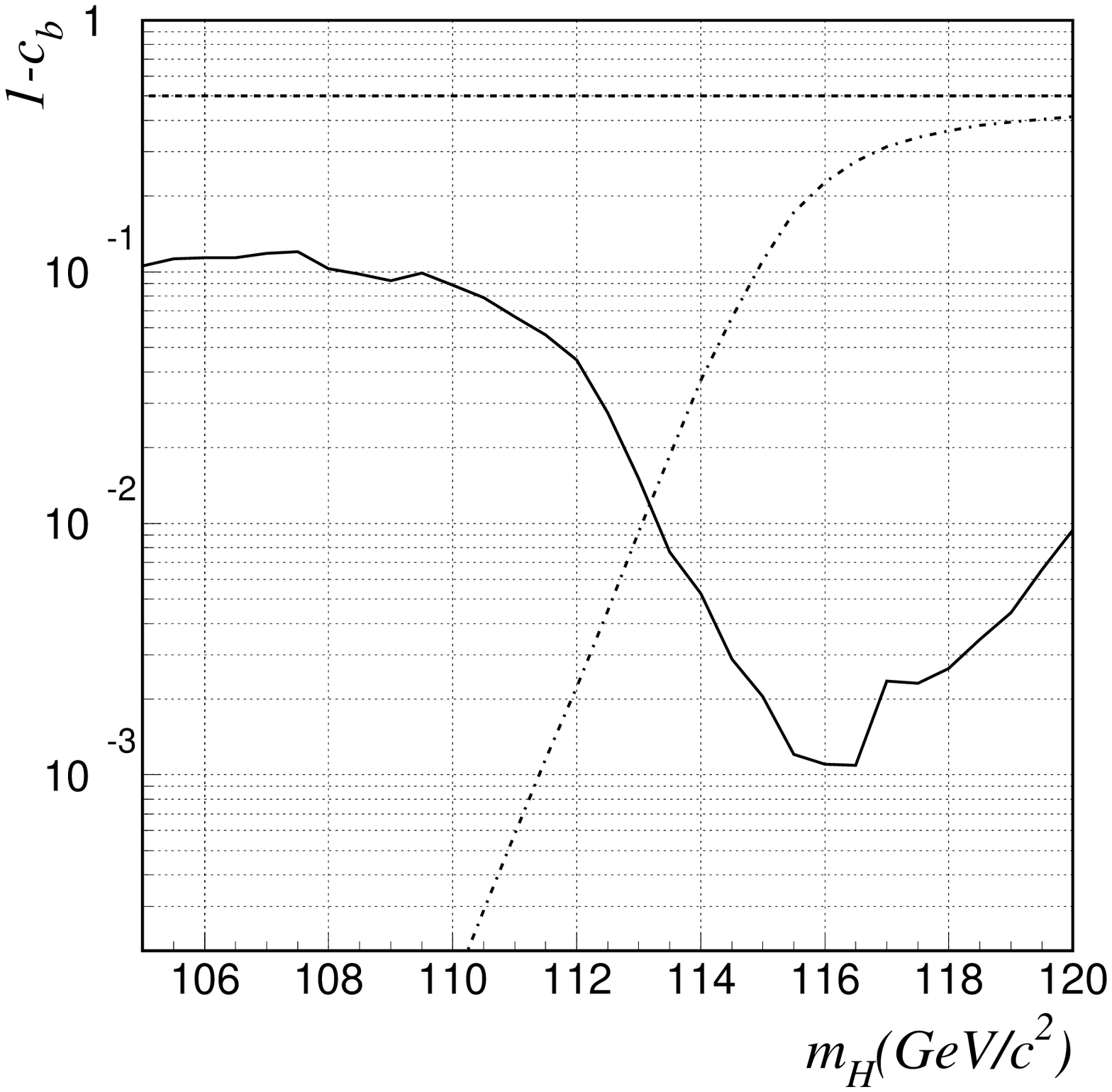}}
\end{picture}
\caption{\capstyl Observed (solid) and expected (dashed) CL curves for
the background hypothesis as a function of the hypothesized
Higgs boson mass for the (a) NN and (b) cut streams. The dash-dotted curves
indicate the location of the median CL for a Higgs boson signal as
a function of the Higgs boson mass.
\label{clb}} 
\end{center}
\end{figure}

Figures~\ref{clb}a (NN) and~\ref{clb}b (cut) show the expected and
observed distributions of $1 - c_b$ as a function of the Higgs boson
mass.  A large deviation from 0.5 can be seen,
consistent with an excess over the
background hypothesis, which is maximal for a Higgs
boson mass of 116\,\Gcs.
The difference between the position of
the likelihood ratio and $1 - c_b$ minima is
due to the inclusion of the expected signal cross section in
the likelihood ratio calculation which, as can be seen in Fig.~\ref{xsect},
decreases rapidly with increasing Higgs boson mass.
The probability of having such a large excess is
$1.5 \times 10^{-3}$ and $1.1 \times 10^{-3}$ for the NN and cut
streams, respectively.
The significance of this excess is $3.0\,\sigma$ and $3.1\,\sigma$
relative to the expected background in the NN and cut
streams\footnote{The LEP Higgs Working Group has adopted a different
convention, using a double-sided Gaussian distribution, which gives
a significance of $3.2\,\sigma$ instead of $3.0\,\sigma$ for the NN
analysis and $3.3\,\sigma$ instead of $3.1\,\sigma$ for the cut
analysis~\cite{pdg}.}.
The expected significance of the excess for a Higgs boson
signal with a mass of 114\,\Gcs\ is shown for each analysis in
Table~\ref{bigtable}.



The Signal Estimator
method~\cite{clse} is used to derive a 95\% CL lower limit on the Standard
Model Higgs boson mass
of 111.1\,\Gcs\
(110.6\,\Gcs) with an expected limit of 114.2\,\Gcs\ (113.8\,\Gcs)
for the NN (cut) stream.

%% file: systematics.tex
\section{Systematic Uncertainties}

Systematic uncertainties in the simulation were
evaluated in a similar manner to Ref.~\cite{Paper1998}
and are
summarized for the expected backgrounds in Table~\ref{bigtable}.
The systematic uncertainty in the expected number of signal events
is typically less than 5\%.

\begin{figure}
\begin{center}
\begin{picture}(170,80)
\put(60,65){\Large \bf (a)}
\put(12,75){\Large ALEPH}
\put(0,0){\epsfxsize=82mm\epsfbox{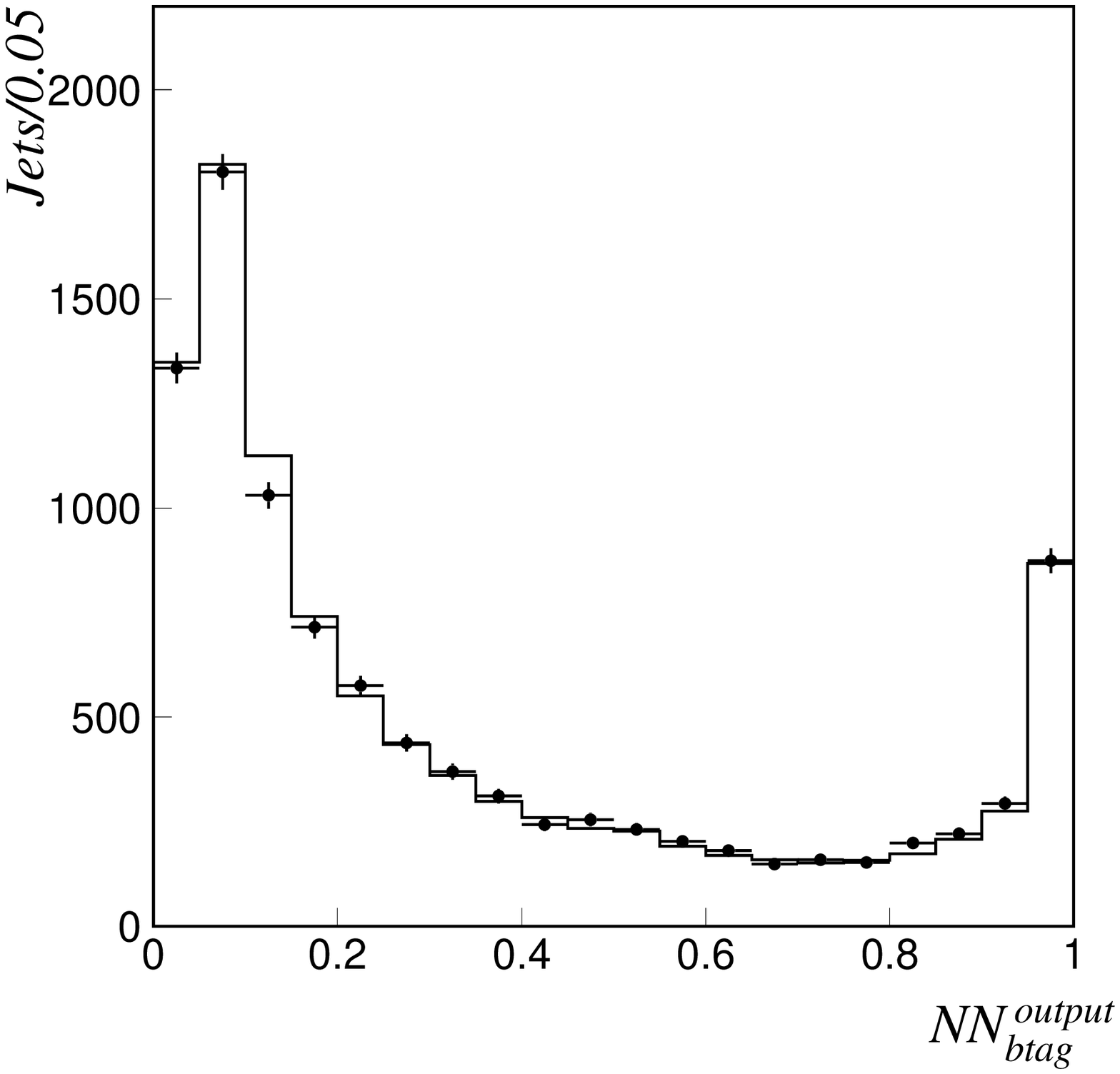}}
\put(145,65){\Large \bf (b)}
\put(97,75){\Large ALEPH}
\put(85,0){\epsfxsize=82mm\epsfbox{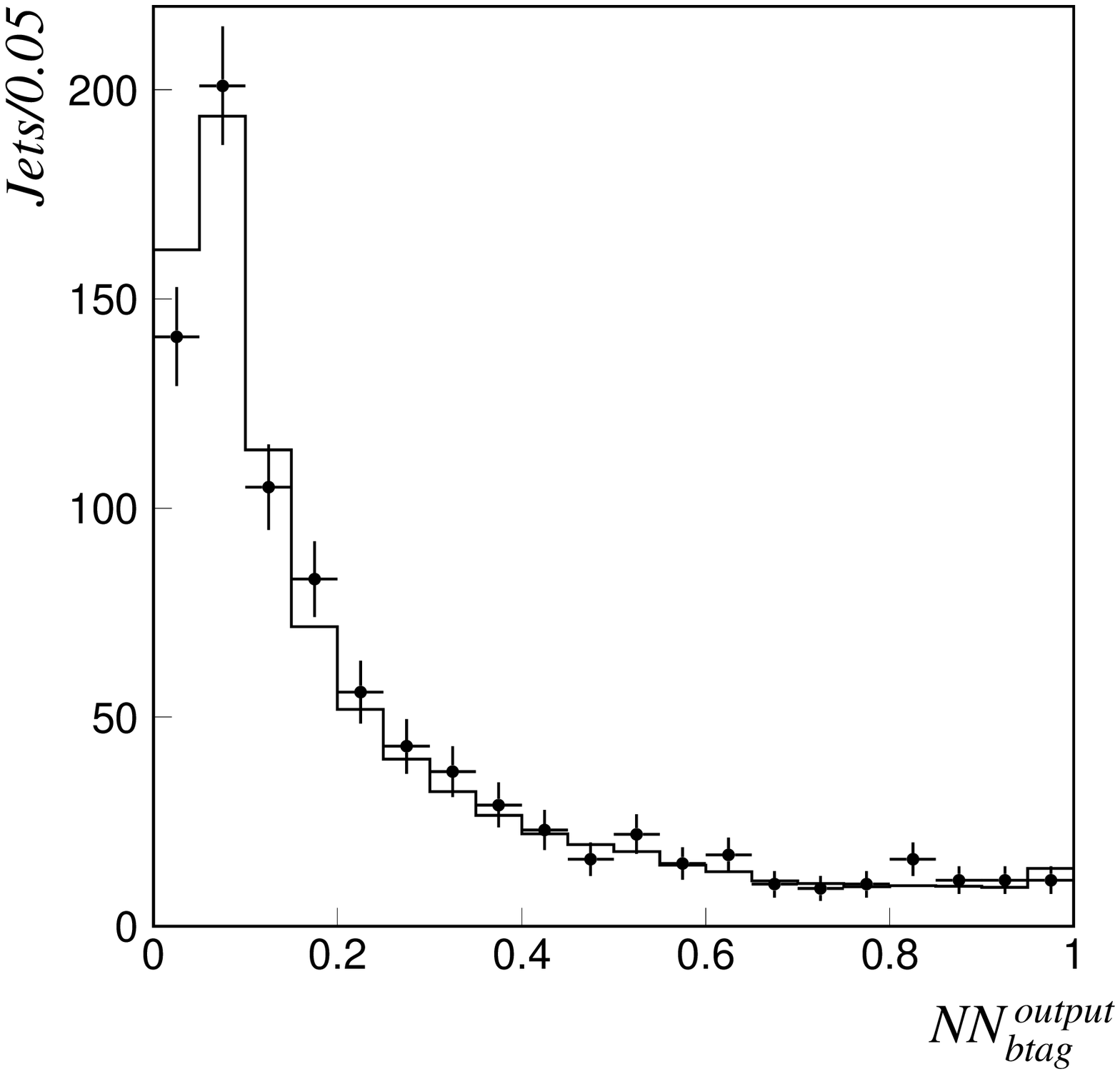}}
\end{picture}
\caption{\capstyl
Distribution of the b tagging neural network output for jets from
(a) radiative Z return events and (b) semileptonic \WW\ events,
in the data
(dots with error bars) and the simulation (histogram).
\label{btagging}} 
\end{center}
\end{figure}

Whenever possible, the systematic uncertainties were
extracted from 3.2\,\invpb\ of data taken at the Z peak during the
same year.
As in previous years, a slight discrepancy between data and
simulation was observed in the impact-parameter-based b
tagging quantities.  To correct for this effect, a smearing of the
track parameters was performed on the simulated events to bring them
into better agreement with the data.  Half of the correction was
taken as a systematic uncertainty.
Figure~\ref{btagging} shows good agreement between the high energy data and
the simulation for both (a) the radiative Z return events, which in 22\%
of the cases contain b quark jets, and (b) the semileptonic \WW\ 
events containing essentially udsc quarks.

The reconstruction of the jet energies and angles has also been studied
with the Z peak data, using the method described in
Ref.~\cite{Paper1998}.  The small differences between data and
simulation are taken into account by smearing the jets in the
simulated events; half of this correction is taken as a systematic
uncertainty.

\begin{figure}
\begin{center}
\begin{picture}(170,80)
\put(52,75){\Large ALEPH}
\put(40,0){\epsfxsize=82mm\epsfbox{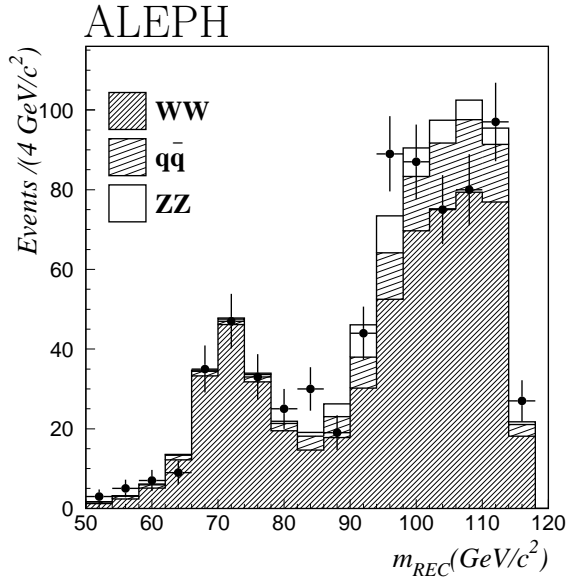}}
\end{picture}
\caption{\capstyl
Reconstructed Higgs boson mass distribution of the
four-jet cut analysis with all kinematic cuts applied and requiring
that no jet has a b tagging neural network output value greater than 0.9.
The data are shown as the dots with
error bars and the simulation, dominated by \WW\ events, is the histogram.
\label{kinmass}} 
\end{center}
\end{figure}

Uncertainties in the distributions of the discriminants used in the
likelihood ratio are dominated in
most analyses by the statistical uncertainties of the simulated samples.

The high mass excess could indicate a possible bias in the mass
reconstruction, not reproduced in the simulation, which would
preferentially select events near the kinematic limit $m_\mathrm{H}
\approx \sqrt{s} - m_\mathrm{Z}$.
The excess is mainly in the 131\,\invpb\ 
of data with centre-of-mass energies greater than 206\,GeV.
To
investigate the possibility a bias, the number of selected events with
reconstructed masses within 5\,\Gcs\ of the kinematic limit were
determined for the four-jet analyses in the 1999 and 2000 data.  For
322\,\invpb\ of data with centre-of-mass energies below 206\,GeV, 11
(4) events were selected with 8.9 (4.6) background events expected
from the NN (cut) analyses.
While for the 131\,\invpb\ of data with $\sqrt{s} > 206$\,GeV, 7
(6) events are selected with 3.5 (1.8) background events expected.
As
the four-jet analyses and the event simulations are unchanged since
1999, there is no evidence of a bias towards the kinematic limit.

To illustrate the reliability of the mass reconstruction,
Fig.~\ref{kinmass} shows the reconstructed Higgs boson mass
distribution from the four-jet cut analysis after all kinematic cuts are
applied and also
requiring that no jet has a b tagging neural network output value
greater than 0.9.  No indication of a non-simulated bias is seen in
the reconstructed mass distribution.  The background is dominated by
\WW\ events; the lower mass peak corresponds to $2m_\mathrm{W} -
m_\mathrm{Z}$, while the broad contribution at high masses is mostly due to
\WW\ events with a wrong pairing assignment. 

As the systematic effects are still under investigation,
the confidence level calculations reported in this letter do not include
the systematic uncertainties.
An estimate of the impact of these uncertainties has been made by
simultaneously increasing the numbers of expected background events by
their errors (Table 1).  The significance of the excess is then reduced
by $0.2\sigma$.  The uncertainty on the distribution of the
discriminants is estimated to have an even smaller effect.

%% file: dissection.tex
\section{Impact of Individual Events}

\begin{figure}
\begin{center}
\begin{picture}(170,80)
\put(60,65){\Large \bf (a)}
\put(12,75){\Large ALEPH}
\put(145,65){\Large \bf (b)}
\put(97,75){\Large ALEPH}
\put(0,0){\epsfxsize=82mm\epsfbox{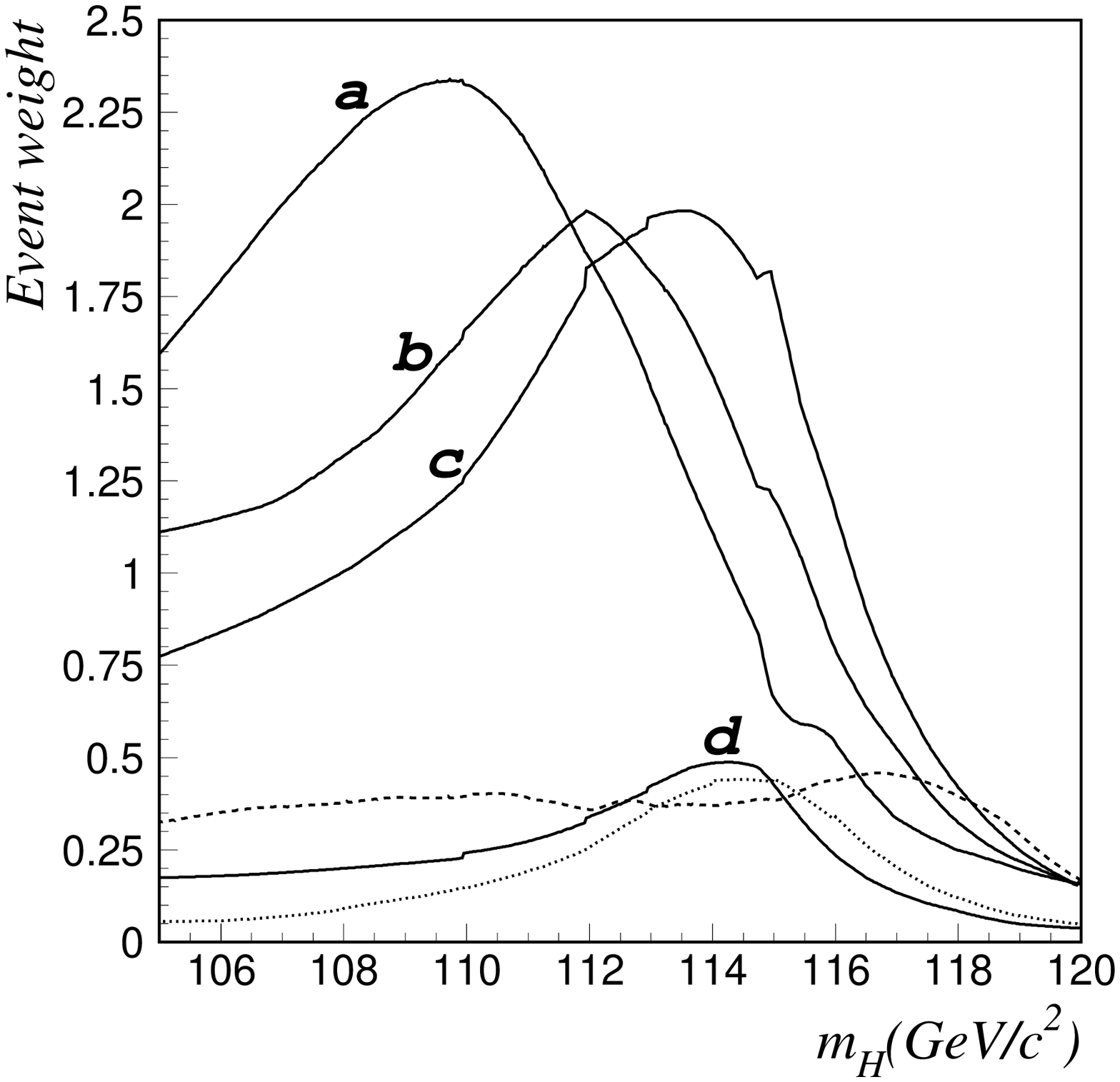}}
\put(85,0){\epsfxsize=82mm\epsfbox{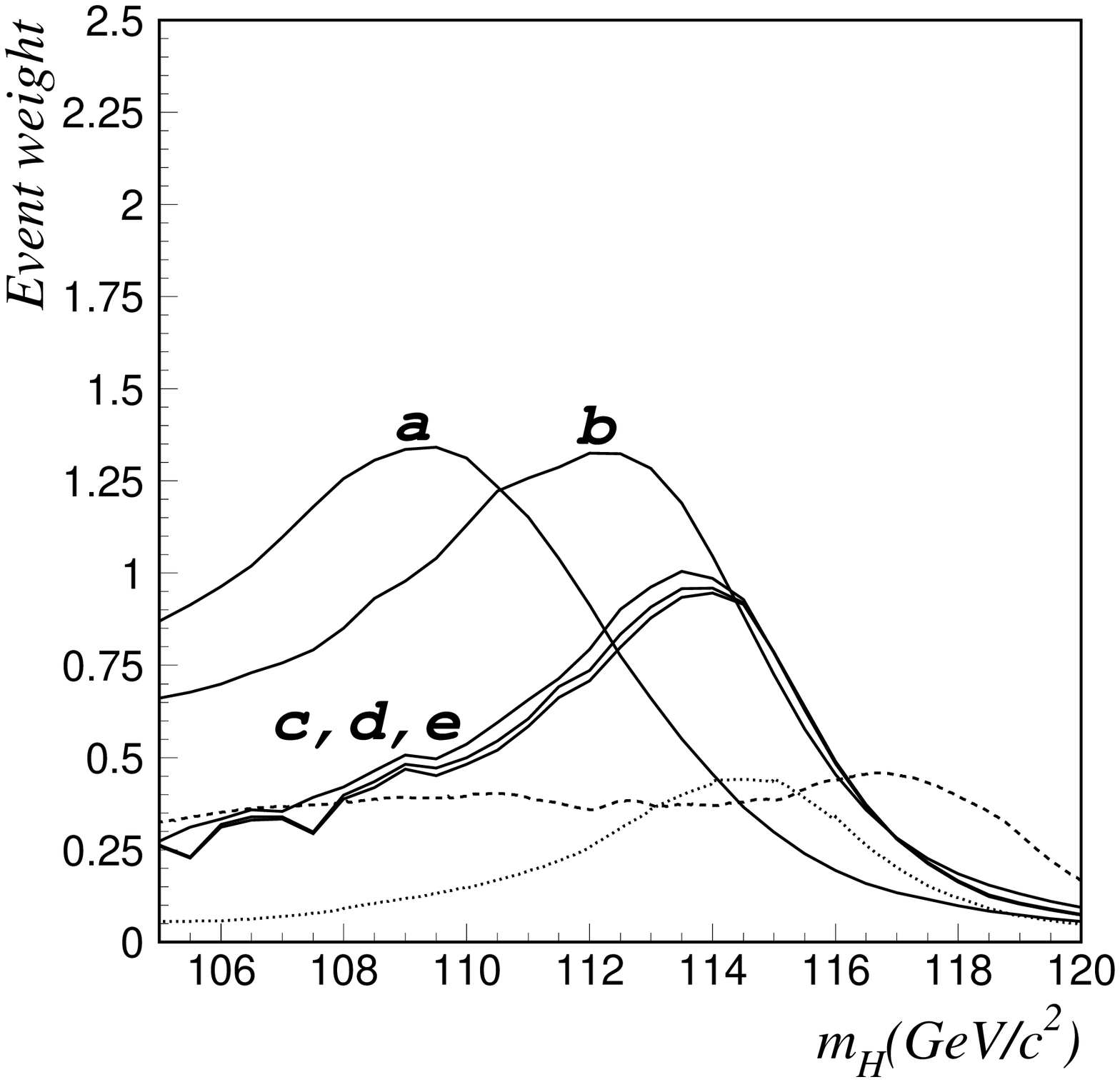}}
\end{picture}
\caption{\capstyl 
Event weight of each candidate as a function of the Higgs boson
mass for the four-jet (solid), lepton pair (dashed) and tau
pair (dotted) final state candidates with a weight larger than
0.4 at a mass of 114\,\Gcs\ in the (a) NN and (b) cut streams.
\label{evolution}} 
\end{center}
\end{figure}

In order to determine the impact of any given candidate event on
the excess, its ``weight'', {\it i.e.}, its contribution to
the logarithm of the likelihood ratio, is calculated as a function
of the Higgs boson mass. 
In Fig.~\ref{evolution}, the event weights are displayed as
a function of the assumed Higgs boson mass for those events with
weights larger than 0.4 at a mass of 114\,\Gcs.

Details of the five four-jet events with weights larger than
0.4 in either the NN or cut analysis are given in Table~\ref{4jetcand}.
All of these events were selected with a centre-of-mass energy
greater than 206\,GeV.
Events $a$, $b$, $c$, and $d$ are retained in both the NN and
cut analyses, while event $e$ has a weight larger than 0.4 only in 
the cut analysis.
The largest contribution to the excess in the NN stream
(Fig.~\ref{evolution}a) comes from three four-jet events ($a$, $b$, and $c$)
which have neural network output values larger than 0.99.
The four-jet cut analysis uses only the reconstructed Higgs boson
mass as discriminant, which causes the three 114\,\Gcs\ events 
($c$, $d$, and $e$) to receive the same weights.
The events with lower reconstructed Higgs boson masses ($a$ and $b$)
have larger weights because they are selected 
in the higher purity 4b final state.

The set of events with weights larger than 0.4 at 114\,\Gcs\ contains
two more events, one in the lepton pair final state and one in
the tau pair final state, which belong to both analysis
streams. No such high weight events are selected by the missing
energy analyses.

\begin{table}[t]
\begin{center}
\begin{tabular}{|c|c|c|c|c|c|c|c|c|}
\hline
\hline
Candidate & Higgs Mass & $m_{12}$ & $m_{34}$ &
\multicolumn{4}{c|}{B tagging} & 4-Jet \\
\cline{5-8}
(Run/Event) & (\Gcs) & (\Gcs)& (\Gcs) & Jet 1 & Jet 2 & Jet 3 & Jet 4 & NN \\
\hline
\hline
$a$ (56698/7455) & 110.0 & 96.3  & 104.9 & 0.999 & 0.836 & 0.999 & 0.214 & 0.999 \\
\hline                                                     
$b$ (56065/3253) & 112.9 & 94.9  & 109.2 & 0.994 & 0.776 & 0.993 & 0.999 & 0.997 \\
\hline                                                     
$c$ (54698/4881) & 114.3 & 101.3 & 104.2 & 0.136 & 0.012 & 0.999 & 0.999 & 0.996 \\
\hline                                                     
$d$ (56366/0955) & 114.5 & 78.8  & 126.9 & 0.238 & 0.052 & 0.998 & 0.948 & 0.935 \\
\hline                                                     
$e$ (55982/6125) & 114.6 & 79.7  & 126.1 & 0.088 & 0.293 & 0.895 & 0.998 & 0.820 \\
\hline                                                             
\hline
\end{tabular}
\end{center}
\caption{\capstyl 
Details of the five four-jet candidates selected with an event weight
greater than 0.4 at a Higgs boson mass of 114\,\Gcs\ in either
the NN or cut streams. The Higgs boson mass is calculated as
$m_{12} + m_{34} - 91.2\,\Gcs$, where jets~3 and~4 are the Higgs boson
jets.
\label{4jetcand}}
\end{table}

The lepton pair final state candidate, recorded at $\sqrt{s}=205$~GeV, is 
shown as the dashed curve in
Figs.~\ref{evolution}a and~\ref{evolution}b. It is reconstructed with a Higgs
boson mass of 118\,\Gcs\ and a b tagging neural network sum of 1.4 for the
two Higgs boson jets.
The invariant mass of the e$^+$e$^-$ pair is 78.8\,\Gcs. The electron
in the event is $6^\circ$ away from one of the hadronic jets. Because the
analysis does not correct for bremsstrahlung photons for electrons within
$10^\circ$ of any jet,
18\,GeV of neutral electromagnetic energy within $2^\circ$ 
of the electron is not considered as bremsstrahlung energy.
If this energy
were added to the leptonic system, its mass would
increase to 93.3\,\Gcs\ and the reconstructed Higgs boson mass
would decrease to 99.5\,\Gcs. 

The \H\tptm\ candidate, recorded at $\sqrt{s} = 208$~GeV, is shown as the 
dotted curve in Figs.~\ref{evolution}a and~\ref{evolution}b. It has a 
reconstructed Higgs boson mass of 115\,\Gcs.
The tau leptons are well isolated and the Higgs boson jets are well
b tagged.
The quality of the kinematic fit is however poor, which is not reflected in 
the event weight since only the reconstructed mass is used as discriminant.

%% file: candidates.tex
\section{High Purity Candidates}

\begin{figure}
\begin{center}
\begin{picture}(170,80)
\put(60,65){\Large \bf (a)}
\put(12,75){\Large ALEPH}
\put(145,65){\Large \bf (b)}
\put(97,75){\Large ALEPH}
\put(0,0){\epsfxsize=82mm\epsfbox{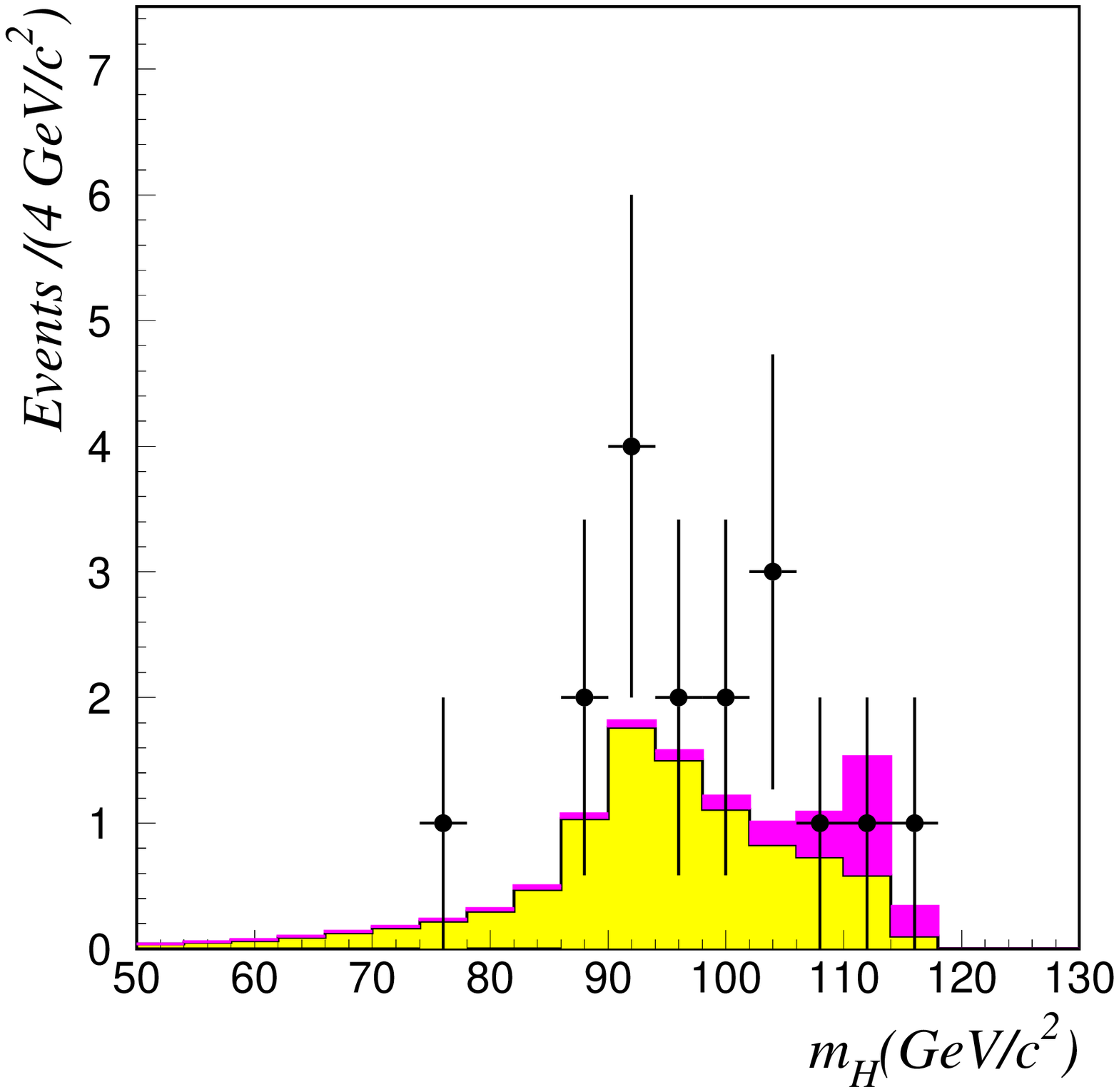}}
\put(85,0){\epsfxsize=82mm\epsfbox{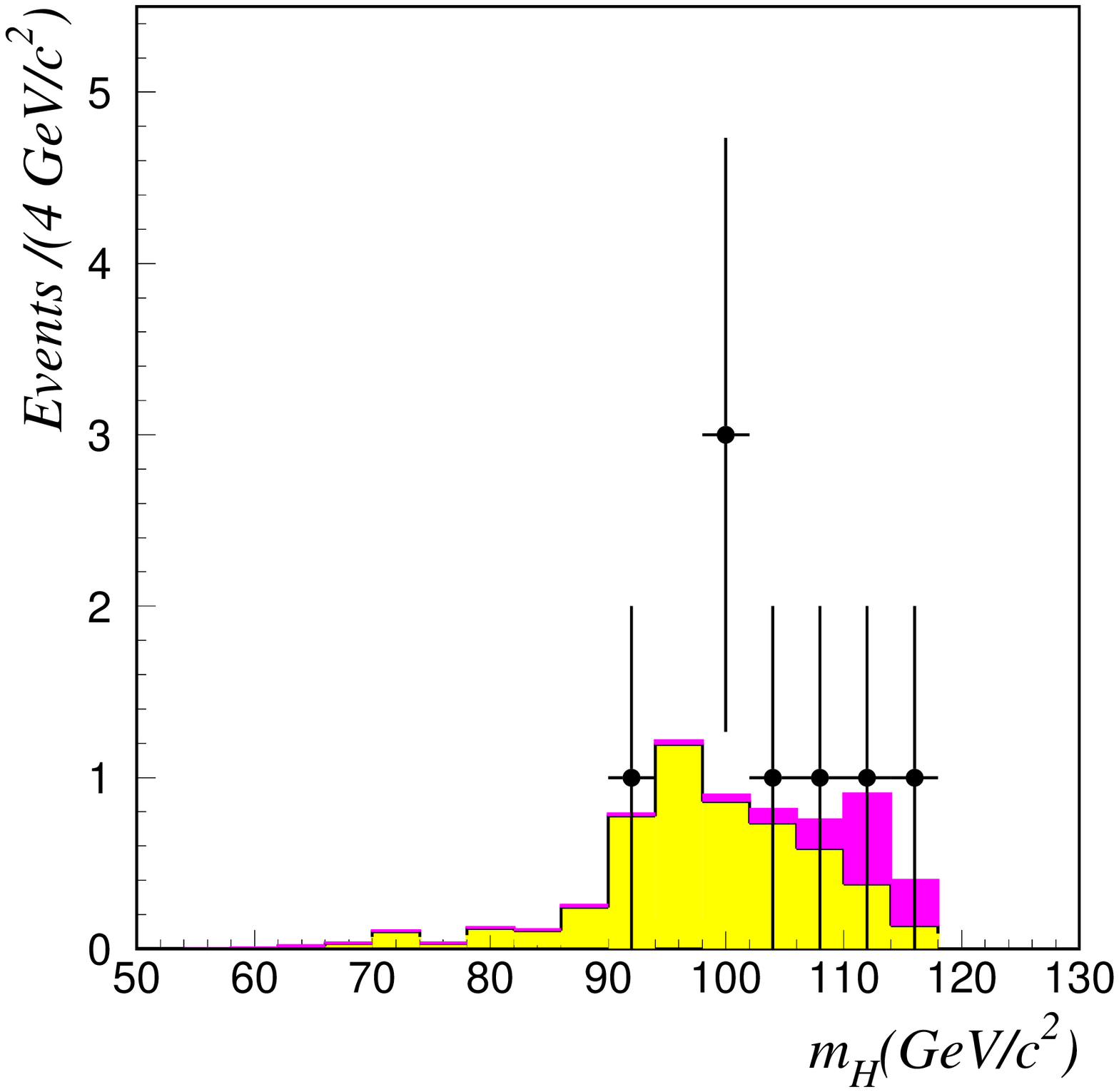}}
\end{picture}
\caption{\capstyl
High purity ($s/b = 1.5$) reconstructed Higgs boson mass
distributions for the (a) NN and (b) cut selections
for the data
(dots with error bars), the expected background (light histogram), and
the expected signal with a Higgs boson mass of 114\,\Gcs\ (dark histogram).
\label{purity}} 
\end{center}
\end{figure}

The stability of the excess can be investigated by increasing the purity
of the event selections. The selection criteria of
all analyses are tightened to give a signal ($m_H = 114\,\Gcs$) to
background ratio ($s/b$) of 1.5 for events with a
reconstructed Higgs boson mass greater
than 109\,\Gcs. Figures~\ref{purity}a and~\ref{purity}b show the high purity
distributions of the reconstructed Higgs boson mass for the NN and cut
streams, respectively.
To obtain a high purity selection in the four-jet NN analysis, the cut
on the neural network output is tightened.
The purity of the four-jet cut
selection is increased
by tightening cuts on the b tagging and the fitted Z boson mass, $m_{12}$.
In the high mass region above 109\,\Gcs, both the
NN and cut streams select the same events, namely the
three four-jet
candidates $a$, $b$, and $c$.
Two of the 114\,\Gcs\ candidates ($d$ and $e$)
significantly affecting the excess in the cut stream
are removed by the tighter cut on $m_{12}$.
In this high mass region,
0.9 (0.6) background events are expected, equally composed of
\qqbar\ and ZZ events, while 1.3 (0.9) signal
events ($m_H = 114\,\Gcs$) are expected for the NN (cut) stream,
all in the four-jet topology.

The three high purity four-jet candidates were reprocessed
with the event reconstruction program
taking into account all of the final detector calibrations
and alignments.
The changes in the reconstructed Higgs boson masses and
neural network values for all three
candidates are insignificant.

\subsection*{Candidate \boldmath{$a$}}

\begin{figure}
\begin{center}
\begin{picture}(170,80)
\put(15,0){\epsfxsize=140mm\epsfbox{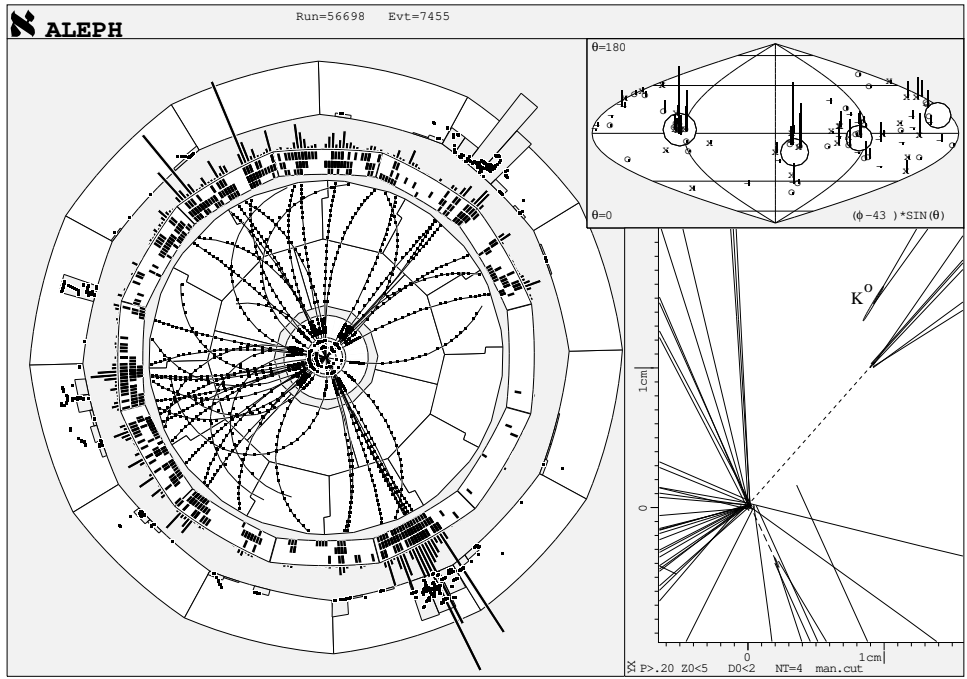}}
\end{picture}
\caption{\capstyl
Four-jet Higgs boson candidate ($a$) with a reconstructed
Higgs boson mass of 110.0\,\Gcs. Three of the four jets are well b tagged.
The event is shown in the view transverse to the beam direction,
the $\theta$-$\phi \sin{\theta}$ view,
and in a closeup of the charged particles in the vertex region.
\label{dali3}} 
\end{center}
\end{figure}

The first high purity candidate ($a$ shown in Fig.~\ref{dali3}),
at a centre-of-mass energy of 206.6\,GeV,
is reconstructed with a Higgs boson mass of 110.0\,\Gcs.
Three of the four jets are well
b~tagged and the event is selected as a 4b event~\cite{Paper1998}
with the sum of the
four b tagging neural network output values equal to 3.05.
The lowest b tagged jet with a value of 0.214 is selected as one of the
Higgs boson jets. The probability for any jet in a 4b event to have
such a low b tagging value is 19\%.

As the event is identified as a 4b event,
any of the six possible pairing combinations can be considered in its
interpretation.
The pairing most compatible
with the ZZ hypothesis, using a fit including the Z boson width and mass
resolutions, gives large fitted Z boson masses of 98.9\,\Gcs\ and
101.6\,\Gcs.

\subsection*{Candidate \boldmath{$b$}}

\begin{figure}
\begin{center}
\begin{picture}(170,80)
\put(15,0){\epsfxsize=140mm\epsfbox{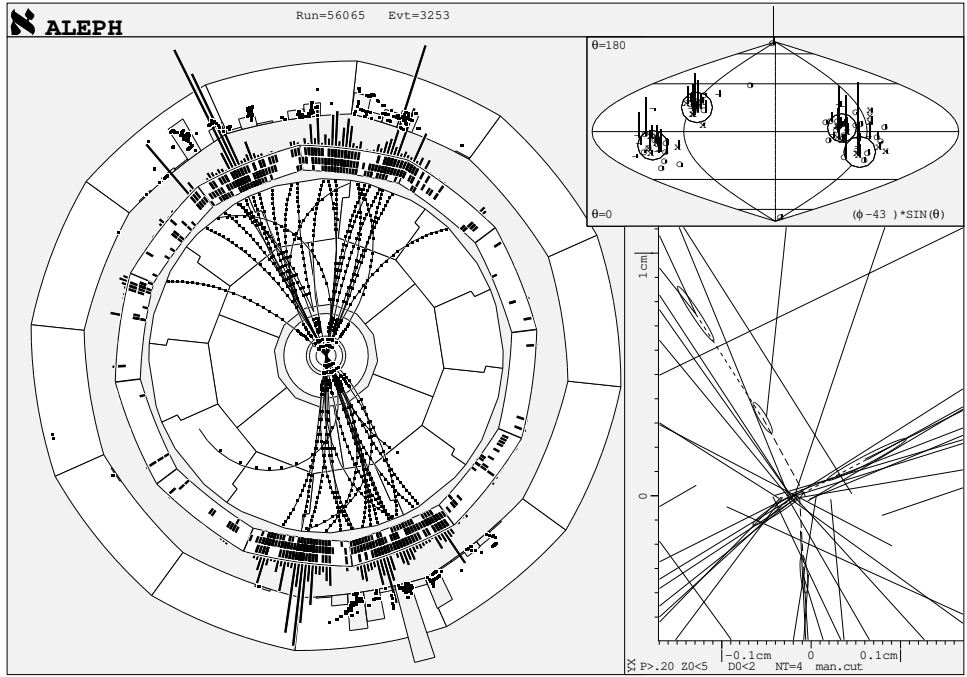}}
\end{picture}
\caption{\capstyl
Four-jet Higgs boson candidate ($b$) with a reconstructed
Higgs boson mass of 112.9\,\Gcs. All four jets are well b tagged.
\label{dali2}} 
\end{center}
\end{figure}

\begin{figure}
\begin{center}
\begin{picture}(170,80)
\put(52,75){\Large ALEPH}
\put(40,0){\epsfxsize=82mm\epsfbox{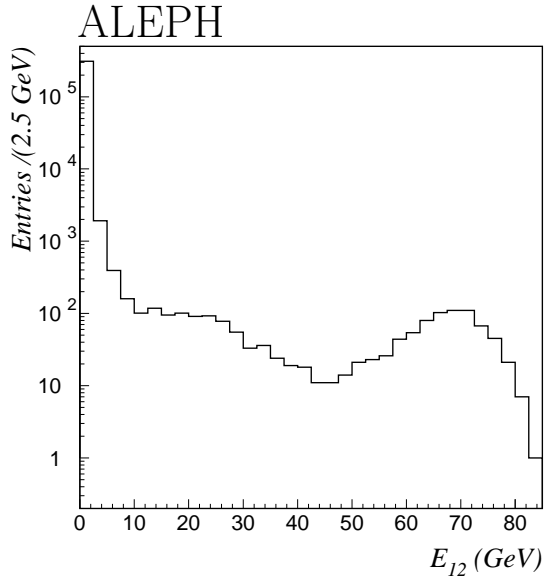}}
\end{picture}
\caption{\capstyl
Energy distribution of the most energetic electromagnetic
particle within $12^\circ$ of the beam in
events triggered at random beam crossings.
\label{e12}} 
\end{center}
\end{figure}

The second high purity candidate ($b$), shown in
Fig.~\ref{dali2}, has a reconstructed Higgs boson mass of 112.9\,\Gcs. 
All four of the jets in the event are well b tagged with a b tagging
neural network
sum of 3.76. The measured visible energy in this event is 252\,GeV,
which is much larger than that allowed by the energy resolution
of about 10\,GeV for
an event with a 
centre-of-mass energy of 206.7\,GeV. A 22\,GeV electromagnetic shower
is detected in the small angle calorimeter (SICAL) in the plane
of the accelerator.
As there is too much reconstructed energy and the momentum
imbalance is in the opposite direction to
the 22\,GeV energy deposit, this shower is most likely a beam-related
particle, unrelated to the rest of the event.
Although the overlapping of such beam-related background is not frequent,
the 22\,GeV of energy is consistent with the observation in
events triggered at random beam crossings,
as can be seen in Fig.~\ref{e12}.

If this low-angle
energy deposit is removed from the event,
the reconstructed Higgs boson mass increases from
112.9\,\Gcs\ to 114.5\,\Gcs. The neural network output for this event
is stable and changes from 0.997 to 0.998.
Because the reconstructed Higgs boson mass shifts closer to the excess,
the significance of the excess would increase by $0.2\,\sigma$
for both the NN  and cut streams.

Even if the 22\,GeV particle is removed from the event, there is still an
energy excess of 23.3\,GeV indicating a mismeasurement of jet
energy.
Such a mismeasurement is often due to fake neutral hadrons,
i.e. hadronic showers which should have been assigned to a
charged particle. This causes
double counting in
the computation of the energy of the jet.
Indeed the detailed inspection of
one of the jets shows that a 13 GeV neutral hadron is likely to have been
misidentified. If this object is removed from the jet and the Higgs boson
mass
recomputed, excluding at the same time the low angle (SICAL) object, a value
of 114.2\,\Gcs\ is obtained. The very small variation in the
reconstructed Higgs boson mass occurs
because the fitted masses depend more strongly on the measured jet directions
than on the jet energies.

As for the first high purity 4b candidate ($a$), the best
background explanation for this event is the ZZ hypothesis with a
different jet pairing. The Z boson masses from a fit
for the most probable ZZ
pairing choice are 94.0\,\Gcs\ and 97.3\,\Gcs.

\subsection*{Candidate \boldmath{$c$}}

\begin{figure}
\begin{center}
\begin{picture}(170,80)
\put(15,0){\epsfxsize=140mm\epsfbox{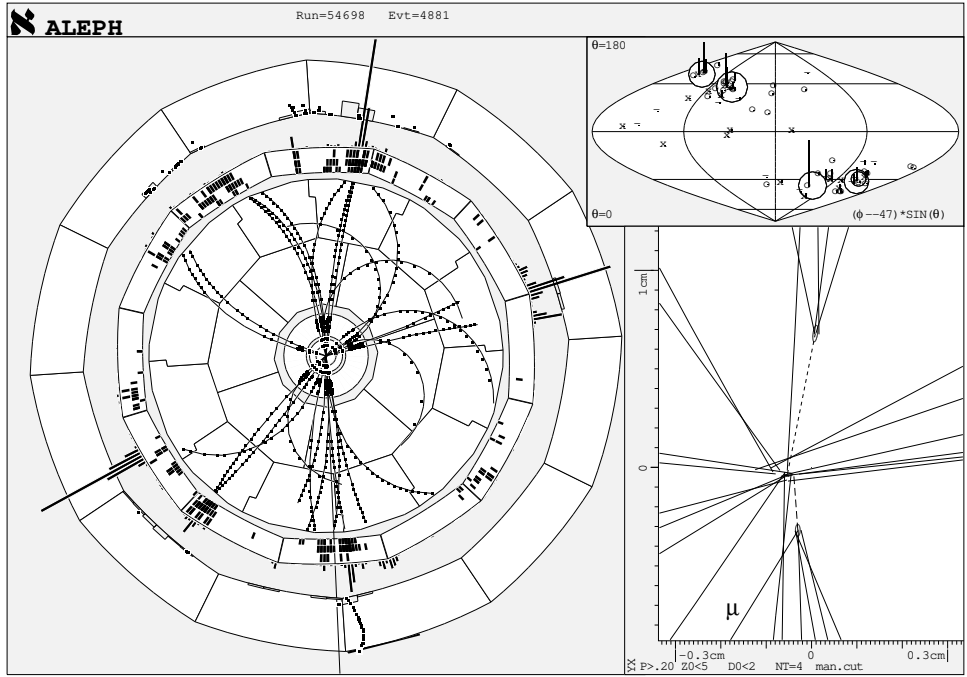}}
\end{picture}
\caption{\capstyl
Four-jet Higgs boson candidate ($c$) with a reconstructed
Higgs boson mass of 114.3\,\Gcs. The two Higgs boson jets are well b tagged.
\label{dali1}} 
\end{center}
\end{figure}

The third high purity four-jet candidate ($c$),
at a centre-of-mass energy of 206.7\,\Gcs,
has a reconstructed Higgs boson mass of 114.3\,\Gcs.
Both of the
Higgs boson jets are very well b tagged with well measured displaced
vertices and b tagging neural net
values of 0.999.
The 13.8\,\Gc\ of missing momentum in the event points to the middle
of the Higgs boson jet containing an identified muon coming
from the secondary vertex, as 
shown in Fig.~\ref{dali1}.
This is a strong indication that, except for the unmeasured neutrino
from the semileptonic b quark decay, the rest of the event is well
measured.
This is also supported by the fact that the measured invariant
mass of the two non b tagged jets is 92.1\,\Gcs, consistent with
a Z boson. The measured invariant mass of the b tagged jets and the
missing momentum is 114.4\,\Gcs,
which renders unlikely the ZZ hypothesis.

Due to the low value of the smallest angle between the four jets
($37^\circ$), the most likely background explanation for this
event is the \bbbar\ggbar\ hypothesis.
The minimum jet-jet angle for the \bbbar\ggbar\ background peaks at low
values with 42\% of the events having angles less than $37^\circ$,
while
11\% of the signal events have
such a low angle.
The two measured jet energies of the non-b jets, 43.5\,GeV and 49.0\,GeV,
are typical, however,
for the decay of a Z boson produced nearly at rest.

%% file: conclusions.tex
\section{Conclusions}

The data collected with the ALEPH detector at
centre-of-mass energies up to 209\,GeV
have been
analysed to search for the Standard Model Higgs boson.
The search was performed using both a neural-network-based
stream and a cut-based stream. Both analysis streams show
a $3\,\sigma$ excess beyond the background expectation,
which is largely
due to candidate events selected in the four-jet analyses
at centre-of-mass energies greater than 206\,GeV.
The
observation is consistent with the production of a
Higgs boson with a mass near 114\,\Gcs.
Reprocessing the most significant candidates using the final
calibration and varying the background expectation by the
systematic uncertainties has shown that these results
are stable.

Results from the four LEP experiments on the search for the
Standard Model Higgs boson were shown at the LEPC meeting
on Nov.\ 3, 2000~\cite{LEPC}.
More data, or results from other experiments, will be needed to determine
whether the observations reported in this letter are the result
of a statistical fluctuation
or the first sign of direct production of the Higgs boson.